\def\yt {{YTe$_{3}$}}
\def\lt {{LaTe$_{3}$}}
\def\ef{{$E_F$}}
\def\a {{\AA$^{-1}$}}
\def\qcdw {$\textbf{q}_{_{\rm{CDW}}}$}
\def\q {$\textbf{q}_1$}
\def\qp {$\textbf{q}_2$}

\documentclass[prx,preprint,amsmath,amssymb,aps,floatfix,longbibliography]{revtex4-2}
\usepackage{float}
\usepackage{wasysym}
\usepackage{graphicx}
\usepackage{dcolumn}
\usepackage{bm}
\usepackage{multirow}
\usepackage{array}
\usepackage{booktabs}
\usepackage{amsmath}
\usepackage[normalem]{ulem}
\usepackage{xr}
\usepackage{physics}
\usepackage[dvipsnames]{xcolor}
\usepackage[section]{placeins}
\usepackage{comment}
\usepackage{hyperref}
\usepackage{cleveref}
\hypersetup{colorlinks=true,citecolor=black, linkcolor=black, filecolor=black, urlcolor=black, pdftitle={Overleaf Example}}
\newcommand\ssA{\textcolor{black}}
\pdfoutput=1 

\begin{document}
	\title {\ssA{Kramers nodal line in the charge density wave state of \yt\ and the influence of twin domains}}


	\author{Shuvam Sarkar$^{1*}$, Joydipto Bhattacharya$^{2,3}$,  Pramod Bhakuni$^{1}$,  Pampa Sadhukhan$^1$, Rajib Batabyal$^1$, \\ Christos D. Malliakas$^4$,  Marco Bianchi$^{5}$, Davide Curcio$^{5}$,   Shubhankar Roy$^6$, Arnab Pariari$^{7}$,  Vasant G. Sathe$^{1}$,  Prabhat Mandal$^{7}$, Mercouri G. Kanatzidis$^{4,8}$, Philip Hofmann$^{5}$,   Aparna Chakrabarti$^{2,3}$, Sudipta Roy Barman$^{1\dagger}$}
	\affiliation{$^1$UGC-DAE Consortium for Scientific Research, Khandwa Road, Indore 452001, Madhya Pradesh,  India}
	\affiliation{$^2$Theory and Simulations Laboratory,  Raja Ramanna Centre for Advanced Technology, Indore 452013, Madhya Pradesh, India}
	\affiliation{$^3$Homi Bhabha National Institute, Training School Complex, Anushakti Nagar, Mumbai  400094, Maharashtra, India}
	\affiliation{$^4$Department of Chemistry, Northwestern University, Evanston, 60208, Illinois, USA}
	\affiliation{$^5$Department of Physics and Astronomy, Interdisciplinary Nanoscience Center (iNANO), Aarhus University, 8000 Aarhus C, Denmark}
	\affiliation{$^6$Vidyasagar Metropolitan College, 39~Sankar Ghosh Lane, Kolkata 700006, India}
	\affiliation{$^7$Saha Institute of Nuclear Physics, HBNI, 1/AF Bidhannagar, Kolkata 700 064, India} 
	\affiliation{$^8$Materials Science Division, Argonne National Laboratory, Lemont, Illinois 60439, USA}

\begin{abstract}
Recent studies have focused on the relationship between charge density wave (CDW) collective electronic ground states and nontrivial topological states. Using angle-resolved photoemission and density functional theory, we establish that \yt\ is a CDW-induced Kramers nodal line (KNL) metal, a newly proposed topological state of matter. \yt\, is a non-magnetic quasi-2D chalcogenide with a CDW wave vector (\qcdw) of 0.2907c$^*$.  Scanning tunneling microscopy and low energy electron diffraction revealed two orthogonal CDW domains, each with a unidirectional CDW and similar \qcdw.   The effective band structure (EBS) computations, using DFT-calculated folded bands, show excellent agreement with ARPES because a realistic x-ray crystal structure and  twin domains are considered in the calculations. The Fermi surface and ARPES intensity plots show weak shadow bands displaced by \qcdw from the main bands. These are linked to CDW modulation, as the EBS calculation confirms. Bilayer split main and shadow bands suggest the existence of crossings, according to theory and experiment. DFT bands, including spin-orbit coupling, indicate a nodal line along the $\Sigma$ line from multiple band crossings perpendicular to the KNL. Additionally, doubly degenerate bands are only found along the KNL at all energies, with some bands dispersing through the Fermi level. 

	\end{abstract}
		\maketitle

\section{Introduction}
Understanding of the coupling between collective electronic ground states such as charge density wave (CDW) or superconductivity and non-trivial topology has become an exciting research frontier in the field of condensed matter physics ~\cite{Sarkar2023,Li2021,Shi2021,Qian2014,Lei2021,Yang2010,Huang2021,Polshyn2022,Zhang2020prb,Mitsuishi2020,Lei2021}. The intricate interplay of CDW and the non-trivial band topology often gives rise to various exotic topological phases such as an axion insulator state~\cite{Shi2021}, Kramers nodal line (KNL) metal~\cite{Sarkar2023}, quantum spin-Hall insulator~\cite{Qian2014},  fractional Chern insulator states~\cite{Polshyn2022}, eightfold fermionic quasiparticle~\cite{Zhang2020prb}
~and  manipulation of topologically protected states~\cite{Mitsuishi2020,Lei2021}, to mention a few. 

CDW is typically observed in layered materials exhibiting quasi-one dimensional (quasi-1D) or quasi-two dimensional (quasi-2D) structures. Due to the associated modification in lattice symmetry, CDW can drive topological phase transitions~\cite{Sarkar2023,Huang2021,Mitsuishi2020,Shi2021}. For instance, we have recently demonstrated that CDW induced inversion symmetry breaking gives rise to a
KNL~\cite{Xie2021}  in the non-magnetic layered rare-earth tritelluride, \lt~\cite{Sarkar2023}. KNLs are new type of two-fold degenerate nodal lines that always connect two time reversal invariant momenta (TRIM) of the Brillouin zone (BZ) and are robust under spin-orbit coupling (SOC)~\cite{Xie2021}. Xie \textit{et al.} have proposed that all non-centrosymmetric achiral crystal symmetries with sizable SOC, when coupled with time reversal (TR)  symmetry, should host this KNL state~\cite{Xie2021}. Besides our recent work on \lt~\cite{Sarkar2023}, experimental evidence of KNL is limited to ruthenium silicides~\cite{Shang2022}
~and SmAlSi~\cite{Zhang2023}.

Yttrium tritelluride (\yt) is another member of the  RTe$_3$ series (R= rare earth)  that is non-magnetic and exhibits an incommensurate CDW state at ambient temperature~\cite{Ru2006,Brouet2008,Yumigeta2021,Malliakas2006}. As in other RTe$_3$ compounds, it has a structure consisting of R-Te1 corrugated blocks sandwiched between the Te2-Te3 bilayers along the long axis ($y$) ~[Fig.~\ref{YTe3_LEED_STM}(a)].  The Te bilayers in \yt\  host the CDW~\cite{Ru2006,Brouet2008}. The CDW transition temperature (T$_{\rm{CDW}}$) of 334~K has been determined from resistivity measurements~\cite{RuThesis}. The magnetization measurements conducted on \yt\ indicate that the diamagnetic susceptibility remains 
~unchanged up to the room temperature~\cite{Ru2006}. The non-magnetic  nature of \yt\ indicates 
 ~the existence of TR symmetry. \yt\ exhibits an orthorhombic structure in the non-CDW state above T$_{\rm{CDW}}$, characterized by the $Cmcm$ space group \cite{Ru2006}. However, in contrast to the other members of RTe$_3$ series (R: La-Tm), whose detailed  x-ray crystallography study in the CDW state is reported in literature~\cite{Malliakas2006}, the structure of \yt\ has not been studied. 
~An angle-resolved photoemission spectroscopy (ARPES) investigation on \yt~\cite{Brouet2008} 
~reported  a $k$-dependent (in-plane) variation of the CDW gap, where the gap decreases as $k_x$ increases. 
~Their study suggests a smaller CDW gap and consequently a smaller gapped region in the Fermi surface (FS)  in \yt\ compared to \lt.    An interesting characteristic of \yt\ is that it exhibits superconductivity  with a   transition temperature of  \AC3 K  with 8\% Pd intercalation, which in turn, inhibits the CDW state~\cite{He2016}. 

The existence of the CDW and resemblance of the  physical properties with \lt\, motivated us to perform a comprehensive study of \yt. At the first step, the structure of \yt\ in the CDW phase was solved by  
~x-ray crystallography. \ssA{Two mutually orthogonal incommensurate CDW domains with similar CDW wave vector (\qcdw) values observed in scanning tunneling microscopy (STM) and low energy electron diffraction (LEED) significantly modifies the ARPES intensity plots and the Fermi surface. Effective band structure (EBS) calculations  based on density functional theory (DFT)  considering the twin domains and using a realistic structure of the CDW state determined by x-ray crystallography  provide excellent agreement with ARPES, which shows faint bilayer split shadow bands that show potential crossing with the main bands. DFT calculations show formation of KNL along the $\Sigma$ line characterized by doubly degenerate bands along the KNL and crossings perpendicular to it.} 

\section{Methods}
\subsection{Experimental:}
	Single crystals of \yt\ with residual resistivity ratio [RRR, $\rho$(300 K)/$\rho$(2 K)]  of $\AC$32 were grown using the tellurium flux method~\cite{Pariari2021}. High-purity Y and Te were mixed in a molar ratio of  \ssA{1:39}. 
	~This mixture was sealed under high vacuum in a crucible and heated at 900$^{\circ}$C for 10 h, and subsequently cooled slowly to 600$^{\circ}$C in 4 days. Excess Te was separated using a high-temperature centrifuge, resulting in gold-colored, plate-like \yt\ crystals. 
	
	Single-crystal x-ray diffraction data for \yt\ were collected at 100~K with the use of graphite-monochromatized MoK$\alpha$ radiation ($\lambda$= 0.71073~\AA) STOE IPDS diffractometer. The collection of intensity data as well as cell refinement and data reduction were carried out with the use of the program X-Area. An analytical absorption correction was performed (X-Shape within X-Area) and the modulated structure was refined with JANA2006~\cite{Petek2014}.
~Atomic coordinates of the atoms in the subcell and initial values of their modulation functions were determined by the charge-flipping method~\cite{Oszlnyi2004,Oszlnyi2004_2}. The distortion (positional or displacement parameter) of a given atomic parameter $x_4$ in the subcell was expressed by a periodic modulation function $p$($x_4$) in a form of a Fourier expansion
$p(k+x_4)$= $\sum_{n=1}^{m}$$A_{sn}$sin$[2\pi\,\overline{q}_n(k+x_4)]$+$\sum_{n=1}^{m}$$A_{cn}$cos$[2\pi\,\overline{q}_n(k+x_4)]$, where $A_{sn}$ is the sinusoidal coefficient of the given Fourier term, $A_{cn}$ the cosine coefficient, $n$ the number of modulation waves used for the refinement and $k$ the lattice translation. $\overline{q}_n$= $\sum_{i=1}^{d}\alpha_{ni}q_i$, where $\alpha_{ni}$ are integer numbers for the linear combination of the incommensurate modulation vectors $q_i$. Satellite reflections of one order were observed and used for the refinement. Consequently, one modulation wave for positional and thermal parameters was used for all atoms. Only the symmetry allowed Fourier terms were refined.

The ARPES measurements were conducted at the SGM3 beamline at the ASTRID2 synchrotron facility~\cite{Hoffmann2004}. FS data at the SGM3 beamline were collected with an energy resolution of 15-20 meV at photon energies (h$\nu$)  of  24 and 28 eV, with an angular resolution of 0.2$^\circ$ (0.008\,\textup{Å}$^{-1}$). These measurements were performed at various temperatures ranging from 45-340~K, and photon energy-dependent studies were conducted using different photon energies in the range of 16 eV to 30 eV. 
~A linearly polarized photon beam in the horizontal plane was incident at an angle of 50$^{\circ}$ with respect to the surface normal, which was oriented along the analyzer axis. The analyzer slit was vertically oriented, resulting in a vertical detection plane, the experimental geometry is similar to what has been used in Ref.~\onlinecite{Sarkar2023}. 
The STM measurements were conducted under a base pressure of 2$\times$10$^{-11}$ mbar employing a variable-temperature STM from Omicron Nanotechnology GmbH in the constant current mode. \ssA{Mechanincally grinded Pt-Ir tips from Unisoku were used and cleaned \textit{in-situ} using voltage pulse method.} LEED was performed using a four grid rear view optics from OCI Vacuum Microengineering. A third order 2D polynomial background function (with 10 coefficients)~\cite{IgorProManualV9} 
~was subtracted from the LEED image to extract the weak CDW related satellite spots. All the measurements were carried out on freshly peeled surfaces under a chamber base pressure of 
~2$\times$10$^{-10}$ mbar.

\subsection{Density functional theory:} 
DFT calculations have been performed using the Vienna Ab-initio Simulation Package (VASP)~\cite{Kresse_1996,Kresse_1999} within the framework of the projector augmented wave method (PAW)~\cite{Kresse_1996,Kresse_1999} to obtain the electronic structure of  \yt. \ssA{The exchange-correlation functional is treated under the generalized gradient approximation~\cite{Perdew}. The energy cut-off is set to 500 eV for the expansion of the plane waves. The convergence criterion for energy in the self-consistent-field cycle and total force tolerance on each atom are taken to be 10$^{-6}$ eV and 0.02 eV/\AA, respectively.  The SOC is employed by a second-variation method as implemented in the VASP code~\cite{Kresse_1999}.}
The  calculations have been performed for a seven-fold approximate structure  with \textit{C2cm} space group (\textit{SG}\,\#40) derived from the experimental atomic positions from the cif file using the PSEUDO program \cite{Capillas2011}. This program displaces the atoms to arrive at the commensurate seven-fold structure with non-centrosymmetric $C2cm$ space group (SG \#40).  This is  discussed further in  section \ref{sec:YT_structure}.

EBS has been computed using Pyprocar code and experimental energy and momentum broadening were convoluted with the unfolded spectral function for comparison with ARPES~\cite{Sarkar2023,Herath2020}. All the DFT bands (and  consequently the EBS)  are rigidly shifted to larger \ssA{binding energy ($E$)} by 0.1 eV  with respect to the \ef\, for comparison with the ARPES data.  VESTA software has been used for crystal structure visualization~\cite{Momma2011}.

	\section{Results and Discussion}
	
	\begin{figure*}[!tb]
	\includegraphics[width=\columnwidth,keepaspectratio,trim={0cm 0cm 0.3cm 0cm },clip]{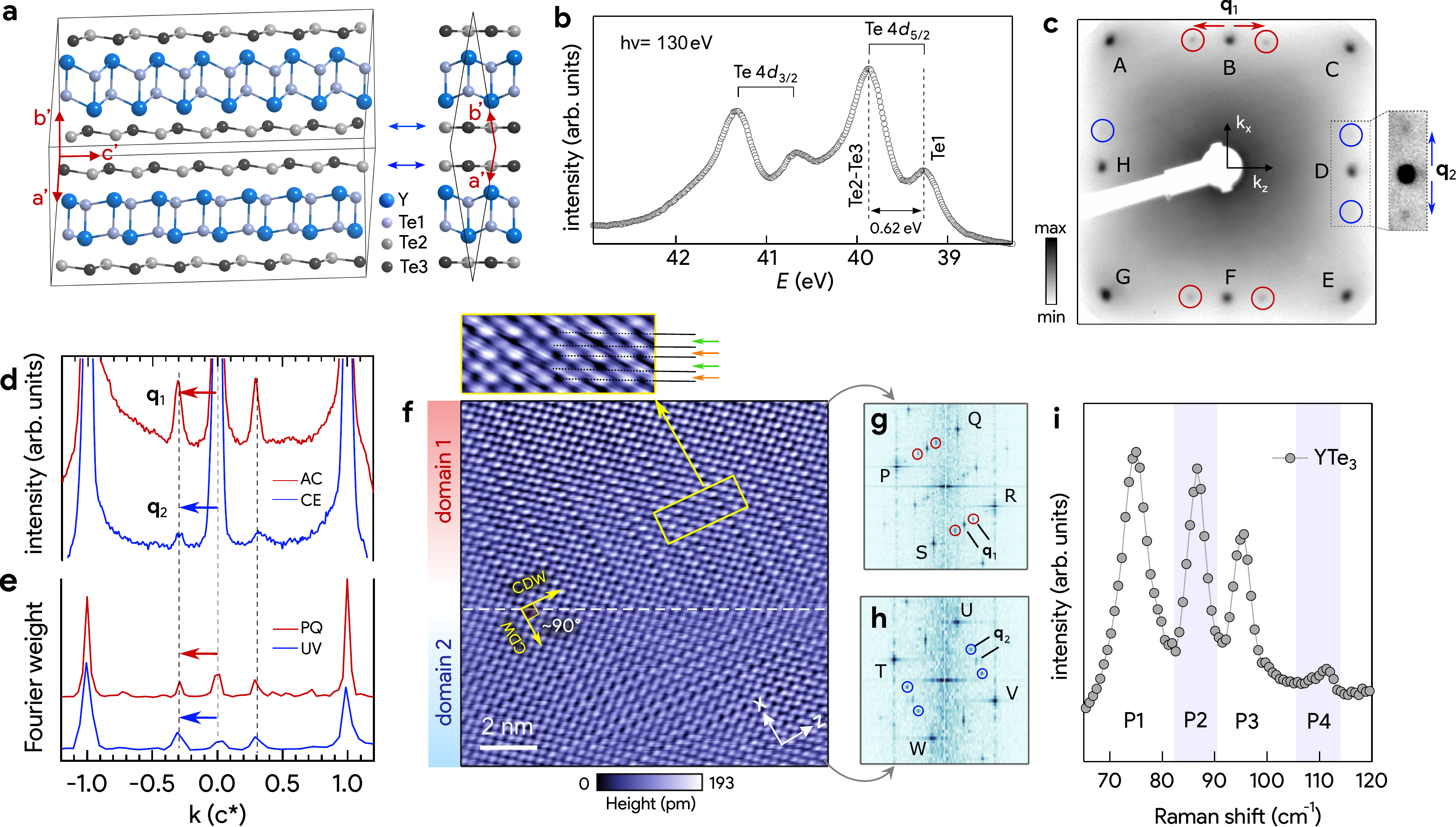}
\caption{(a) The primitive unit cell  
	~of the seven-fold structure  (1$\times$1$\times$7) for   \yt\ in the CDW state comprising of 56 atoms with  \q= $\frac{2}{7}$$c^*$ ($c^*$ is the reciprocal lattice vector along $k_z$ in
	the non-CDW state)
	~\textit{left}: viewed perpendicular  and \textit{right}: parallel  to  the \textbf{c$^{\prime}$} direction. The lattice constants  are $a^{\prime}$= $b^{\prime}$= 12.781 \AA, $c^{\prime}$= 29.999~\AA~ with $\alpha^{\prime}$= $\beta^{\prime}$= 90$^{\circ}$, and $\gamma^{\prime}$= 160.76$^{\circ}$.  (b) Te 4$d$ core level spectrum measured using  h$\nu$= 130 eV. 
	~(c) The low energy electron diffraction (LEED) pattern  measured with 39  eV primary electron beam energy,  the CDW satellite spots are encircled. The color bar is in log-scale to enhance the weak satellite spots. (d) The intensity line profiles along AC and CE in \ssA{panel \textbf{c}}. The red and blue arrows indicate the CDW wavevectors \q\ and \qp\ along $k_z$ and $k_x$, respectively. (e) The intensity line profiles along PQ in  panel \textbf{g} and UV in panel \textbf{h}.	(f) A  scanning tunneling microscopy (STM) topography image (12.8$\times$12.8 nm$^2$) in the CDW state of \yt, with bias voltage of  0.2 V and a tunneling current of  0.37 nA. The color scale in the bottom is shown in picometers (pm). A zoomed image of the	 enclosed area inside the yellow box is shown in the inset.  Fourier transforms (FT) of the (g) top (above the white dashed line that indicates the domain boundary) and (h) bottom (below the white dashed line) part of the STM image. The CDW related satellite spots are encircled. Both LEED and STM measurements were carried out at room temperature (RT). (i) The Raman spectrum (gray circles) of \yt\ measured at  90 K shows four peaks indicated as P1, P2, P3 and P4, where the P2 and P4 peaks (highlighted by the light-blue rectangles) contain the \textit{B}$_{\rm 1}$ symmetry mode.}
\label{YTe3_LEED_STM}	
\end{figure*}

\subsection{Crystal structure  of \yt\ in the CDW state}
\label{sec:YT_structure}
  
X-ray crystallography data  show that \yt\ has an orthorhombic structure. The lattice constants and other crystal data are presented in Table~S1 of the
~Supplementary Material (SM)~\cite{supp}. We find that \yt\ hosts a unidirectional incommensurate CDW with the \qcdw= 0.2907c$^*$  and the superspace group is determined to be  $C2cm$(00$\gamma$), similar to the other members of the RTe$_3$ series~\cite{Malliakas2006}. $C2cm$ is the basic space group of the $C2cm$(00$\gamma$) superspace group and $\gamma$ represents the $z$ component of \qcdw.   

An incommensurate structure is generally approximated as a commensurate structure with a large unit cell such that its \qcdw\ is close to  the incommensurate value. However, the latter becomes an accurate representation of the incommensurate structure if it is based on the atom positions determined by  x-ray crystallography and the \qcdw\ value  is within the experimental accuracy~\cite{Janssen2006,van2007incommensurate}. By utilizing the continued fraction method~\cite{Wyman1985} to find a rational fraction that could represent \qcdw, we arrive at a fiftyfive-fold structure (1$\times$1$\times$55 supercell) with \textit{C2cm} space group with \qcdw\ = 16/55c$^*$ = 0.2909c$^*$ that matches the experimental value of 0.2907(4)c$^*$ within its accuracy. It has 440 atoms in the unit cell with positions almost coinciding with those given by x-ray crystallography. 
~A comparison   of the Te2 and Te3 atom positions (green dots) with the experimental positions (open circles) show no perceptible deviation [Fig.~S1(a)
~of SM~\cite{supp}]. This is also supported by a small average displacement ($u$= 0.001~\AA) of the atoms in the unit cell from the experimentally determined positions, 
~as obtained from the PSEUDO program~\cite{Capillas2011}.

However, because of the large size of the fiftyfive-fold unit cell, DFT calculations \ssA{turned out to be highly resource intensive}. 
~So, a  smaller seven-fold unit cell with 56 atoms is derived, importantly with all symmetries of the fiftyfive-fold structure preserved i.e., with the same space group [Fig.~\ref{YTe3_LEED_STM}(a)]. The corresponding  \qcdw\ value is 2/7$c^*$ (= 0.2857$c^*$), which is an 1.8\% deviation from the experimental value, with the Te atoms  (orange dots) showing small  deviations  from the experimental positions (open circles) [Fig.~S1(b)
~of SM~\cite{supp}].  Figure~\ref{YTe3_LEED_STM}(a) shows that \yt\, is made up  of two main structural units:  the Te2-Te3   bilayer that hosts the CDW and the \ssA{Y-Te1} corrugated slab. The Te bilayer, highlighted by blue double-sided arrows,  is weakly coupled by van der Waals interaction. 

Te 4$d$  core-level photoemission spectrum  of \yt\  in Fig.~\ref{YTe3_LEED_STM}(b) comprises of   4$d_{5/2}$ and  4$d_{3/2}$ peaks separated by the spin-orbit splitting of \AC1.5 eV. 
~Note that each of these peaks exhibit two components that are separated by \AC0.6 eV.  These components are the signature of  different valency of  Te in  the Y-Te1 slab and the Te2-Te3 layer due to transfer of electronic charge from the former to the latter. \ssA{ A nearly similar splitting 
	was  observed in \lt\ 4$d$~\cite{Sarkar2023} and 3$d$~\cite{SarkarAIP2020} spectra. This was attributed to the difference in valencies of  Te1 compared to Te2 (and Te3), as supported by the DFT calculations.} 

\subsection{Twin domains on the \yt\ surface}
\label{sec:YT_LEED}

In Fig.~\ref{YTe3_LEED_STM}(c), the LEED pattern of  the \yt\ surface  in the CDW state at room temperature (the CDW transition temperature $T_{\rm CDW}$ being 334~K~\cite{RuThesis})  displays sharp main spots labeled as A-H.  The satellite spots   related to the CDW modulation along  the $k_z$ direction  are highlighted by red circles. Interesting to note are the relatively weaker satellite spots  in the $k_x$ direction (blue circles). These are unambiguously visible  after performing a background subtraction, 
~as shown in the inset on the right side.  The CDW modulation vectors \q\ along $k_z$ and \qp\ along $k_x$ have been determined from the LEED intensity profiles by measuring the distance between the satellite and the  main peaks. In Fig.~\ref{YTe3_LEED_STM}(d), the intensity line profiles  measured along AC and CE demonstrate that both \q\ and \qp\ have similar value of (0.3$\pm$0.01)$\textbf{c}^*$, which is in agreement with that determined by x-ray crystallography. Note that for  RTe$_3$ with R= Tb-Tm, coexisting
~bidirectional CDW has been observed~\cite{RuPRB2008,Yumigeta2021,Fang2007}, \ssA{and the magnitudes of \qcdw\  in the two  mutually perpendicular directions are different (e.g.,  (2/7)\textbf{c}$^*$ and (1/3)\textbf{a}$^*$  for ErTe$_3$ \cite{Brouet2008,Moore2010}). On the other hand, the \qcdw\ values for the two twin domains are similar here for \yt.}   

These twin domains are directly observed in the STM topography image with atomic resolution, where  the domain boundary is  indicated by a dashed white line  [Fig.~\ref{YTe3_LEED_STM}(f)]. Satellite spots related to the CDW are also visible in the Fourier transforms (FT)  of each domain along $k_z$ and $k_x$ for  domain \textit{1} (top) and domain \textit{2} (bottom), respectively [Figs.~\ref{YTe3_LEED_STM}(g,h)]. In Fig.~\ref{YTe3_LEED_STM}(e), the \textbf{q} values for the two domains from the line profiles   measured along PQ [Fig.~\ref{YTe3_LEED_STM}(g)] and UV [Fig.~\ref{YTe3_LEED_STM}(h)]  are \q= (0.29 $\pm$ 0.03)\textbf{c}* and \qp= (0.3 $\pm$ 0.05)\textbf{c}*, which are equal within the error bar and are in excellent agreement with the values obtained from LEED. \ssA{It was also noted from STM topographies performed at different sample locations 
	~that domain \textit{1} is more prevalent compared to domain \textit{2}, a finding corroborated by the lower intensity of the LEED spots corresponding to domain \textit{2} [Figs.~\ref{YTe3_LEED_STM}(c)]. }

In Fig.~\ref{YTe3_LEED_STM}(f), atomic resolution enables direct observation of the  CDW in the top Te layer.  A zoomed region in the yellow rectangle shows that the average positions of  the neighboring Te chains, depicted by the black dashed lines, are not equidistant (orange and green arrows), signifying the breaking of $M_x$ mirror symmetry and, consequently, the inversion symmetry as well. The non-centrosymmetry  is  also demonstrated by the  Raman spectrum in Fig.~\ref{YTe3_LEED_STM}(i) through occurrence of  P2 and P4 peaks of  $B_{\rm 1}$ symmetry mode, which is an irreducible representation of \ssA{achiral} C$_{2v}$ point group. These modes were also observed in  \lt~\cite{Sarkar2023,Lavagnini2008}.   The mirror symmetry however remains intact in the non-CDW state that has a centrosymmetric $Cmcm$ space group~\cite{Ru2006,RuThesis}.

\begin{figure*}[!tb]
	\includegraphics[width=0.8\columnwidth,keepaspectratio,trim={0 0cm 0 0 },clip]{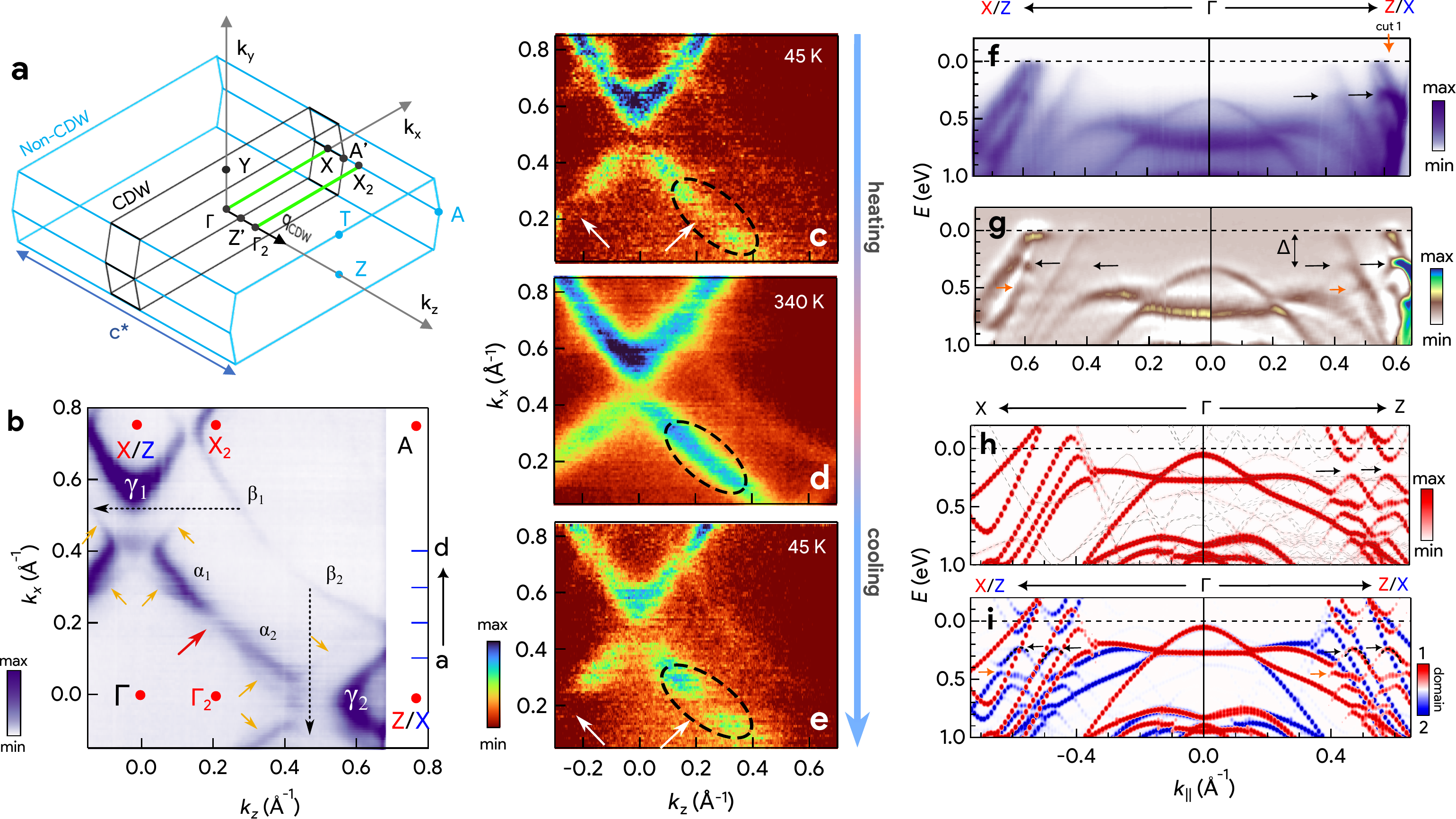}
\caption{ (a) The Brillouin zone (BZ) of the seven-fold structure of the CDW-state  of \yt\ (black lines), where the ordering of the conventional lattice is $a_c^{\prime}<b_c^{\prime}$~\cite{Setyawan2010}  within the BZ of the non-CDW state (cyan). The   high symmetry points~\cite{Setyawan2010} are indicated in the respective colors. $\Gamma$, $Y$ and $X$ points coincide for both.  The $\Sigma$ line represented by $\Gamma X$ (and  $\Gamma_2 X_2$  in the 2$^{\rm nd}$ CDW BZ) is shown by green color.  (b) The Fermi surface (FS) of \yt\ in the CDW state measured  at 45 K with h$\nu$= 24 eV. In panel \textbf{b}, the horizontal (vertical) black dashed arrow of length  \q\ (\qp), connects a shadow branch with the main branch of the FS. The main branches, $\alpha$, $\beta$, and $\gamma$, are denoted with subscripts 1 and 2 to denote the branches of  domains \textit{1} and \textit{2},  respectively. The gapped region in the $\alpha$ branch around $k_z$=  $k_x$= 0.2 \a\ is highlighted by a red arrow.  The yellow arrows indicate the shadow branches. The high symmetry points of the BZ  are indicated by red and blue colors, denoting  domain \textit{1} and \textit{2}, respectively. The color scale is shown on the left. Blue lines on the right axis indicate the directions of
	~the $E(k_z)$ cuts in Figs.~S2(a-d) of SM~\cite{supp}. 
~Temperature dependent measurement of the FS  with h$\nu$= 28 eV at  (c) 45~K (d) 340~K and subsequently cooled to (e) 45~K. Black dashed ovals highlight the region of interest. The color scale for panels \textbf{c}-\textbf{e} is shown on the left of panel \textbf{e}. (f) $E$($k_z$) band dispersion along $X\Gamma Z$ direction. Band bending  due to CDW gap are indicated by the black arrows.  (g) The curvature plot of  panel \textbf{f}, where the band bending is more prominent (black arrows). $\Delta$ is related to the CDW gap. (h) The folded band structure (dashed gray curves) for single domain of  \yt\  in the seven-fold structure along X$\Gamma$Z   including spin orbit coupling (SOC) is overlaid on the effective band structure (EBS) \ssA{calculated from it.} (i) The EBSs of domain \textit{1} (red) and domain \textit{2} (blue) overlaid on each other. The black arrows indicate the band bending related to the CDW gap.}  
\label{YTe3_FS}	
\end{figure*}

\subsection{Shadow branches in the twin domain modified Fermi surface}
\label{sec:YT_FS_LT}

The Fermi surface  in the   CDW state  of \yt\  measured by ARPES  comprises two diamond-shaped sheets centered around $\Gamma$ and two smaller oval pockets -- $\gamma_1$ and $\gamma_2$ -- located near the $X$ and $Z$ points, respectively; these high symmetry points are indicated by red color [Fig.~\ref{YTe3_FS}(b)]. The  high symmetry points are also shown in the Brillouin zone (BZ) (black lines) in Fig.~\ref{YTe3_FS}(a), it is  inscribed within the non-CDW BZ (cyan). 
 ~A signature of a gapped region in the FS around the $\Gamma Z$ direction  was  observed previously for other RTe$_3$ compounds~\cite{Brouet2008, Sarkar2023}. Curiously, such a gapped region seems to be absent in  Fig.~\ref{YTe3_FS}(b). The FS rather seems to nearly resemble \ssA{that calculated for the non-CDW state of RTe$_3$~\cite{Sarkar2023, Brouet2008}, except for 
~an important difference:}  occurrence of a narrow gapped region,  highlighted by the red arrow at ($k_x$, $k_z$) = (0.2, 0.2) \a\  that splits the inner diamond-shaped sheet ($\alpha$). We henceforth refer to the upper (lower) part as $\alpha_1$ ($\alpha_2$), as shown in  Fig.~\ref{YTe3_FS}(b). $\alpha_2$ resembles  $\alpha_1$ rotated by 90$^{\circ}$. A similar replica of  the  $\beta_1$ (upper part of the outer sheet)   branch is named as $\beta_2$ (lower part).  $\gamma_2$ that is observed around \textit{Z} in the lower right corner, is also  a  replica of the $\gamma_1$ pocket around \textit{X} (upper left corner). 

The appearance of these replicas in the FS  rotated by 90$^{\circ}$ can be explained by the presence of  the twin domains (\textit{1} and \textit{2}) since the ARPES signal with a photon beam spot size of (200$\times$100)~$\mu$m$^2$ probes both  the domains simultaneously.  The intensity of the FS branches related to domain \textit{2} is less  e.g., compare the intensities of the $\alpha_2$ sheet and the $\alpha_1$ sheet.  This indicates that the contribution of domain \textit{1} is greater than that of domain \textit{2}, which is consistent with the LEED and STM results \ssA{discussed in the previous section}. The high symmetry points  of  domain \textit{1} (related to \q) and domain \textit{2} (related to \qp) are shown by  red and blue colors, respectively in Fig.~\ref{YTe3_FS}(b).  

To examine whether the narrow gapped region (red arrow) mentioned above might be a signature of the CDW gap, we have measured the FS in both CDW and the non-CDW states. 
~Interestingly, in contrast to the FS in the CDW state at 45 K [Fig.~\ref{YTe3_FS}(c)],   as the temperature is raised above   $T_{\rm CDW}$ [Fig.~\ref{YTe3_FS}(d)], the $\alpha$ sheet is no longer gapped  at 340~K  (black dashed oval), indicating CDW melting. Subsequently, as the temperature is lowered back to 45 K, the gap  becomes visible again [white arrows in Fig.~\ref{YTe3_FS}(e)]. Thus temperature dependent FS measurement establishes  that the narrow gapped region in the $\alpha$ sheet corresponds with the CDW gap in \yt. Additionally,  from  a comparison of the ARPES bands and the calculated EBS considering both domains, we show  how the combined effect of CDW gap and twin domains gives rise to the narrow gapped region in the FS in Discussion~A
~of the SM~\cite{supp}. 

\ssA{Thus, unlike the case of  RTe$_3$ single crystals (as in Ref.~\onlinecite{Sarkar2023}), the FS of our YTe$_3$ crystals is significantly modified by the presence of twin domains.} 
~In spite of this,   shadow branches that are related to the CDW modulation are observed, as indicated by dark yellow arrows in Fig.~\ref{YTe3_FS}(b). These are separated from the main branches by \qcdw (black dashed arrows). The shadow branches are observed for both the domains; see for example, the vertical and horizontal black dashed arrows of length \qcdw.  

\subsection{ARPES and effective band structure along X$\Gamma$Z}
 The ARPES intensity plot of \yt\ along X$\Gamma$Z high-symmetry direction in Fig.~\ref{YTe3_FS}(f)  shows that the bands towards the $\Gamma Z$ ($\Gamma X$) direction of domain \textit{1} overlap with those of the $\Gamma X$ ($\Gamma Z$) direction of domain \textit{2}. 
 ~Therefore, the intensity plot 
 ~seems to be essentially identical in these two directions, with bands crossing the \ef\ in both directions. Additionally, in both the directions, some bands are bent away from the \ef\ [black arrows] due to the hybridization of the main band  and shadow band, resulting in formation of the CDW gap. 
~In Fig.~\ref{YTe3_FS}(g), these hybridized bands with enhanced clarity are shown by the black and orange arrows around $E$= 0.3 eV and 0.5 eV, respectively. The lower limit of the CDW gap defined from the \ef\ to the band maximum  [$\Delta$], is estimated to be   0.29$\pm$0.02 eV. This value is  similar in both directions.

In Fig.~\ref{YTe3_FS}(h), the DFT calculated bands of \yt\ along $X\Gamma Z$ for the  seven-fold structure   show multiple folded bands along both $\Gamma Z$ and $\Gamma X$ directions (gray dashed curves), which are difficult to compare with the  ARPES intensity plots [Figs.~\ref{YTe3_FS}(f,g)].  So, we have  determined the EBS from the DFT bands by band unfolding, as shown by the red markers~\cite{Sarkar2023,Ku2010,Allen2013}. The EBS shows reasonable resemblance with ARPES: The CDW gap along $\Gamma Z$  is evident in the EBS, as marked by the black arrows in Fig.~\ref{YTe3_FS}(h).   Along $\Gamma X$, two parabolic bands (related to the $\gamma$ pocket) centered around \textit{X} cross the \ef, along with two additional bands (related to the $\alpha$ sheet) that approach the \ef. 

The agreement between theory [Fig.~\ref{YTe3_FS}(i)] and experiment [Fig.~\ref{YTe3_FS}(g)] is excellent
~when the influence of the twin domains is accounted for by superimposing the EBSs calculated along  $\Gamma Z$ and $\Gamma X$, where the red and blue bands represent the  domain \textit{1} and \textit{2} bands, respectively.  For $k_{||}$$>$0, the bands involved in the CDW gap formation are from  domain~\textit{1}, while the bands that cross \ef\  are from domain~\textit{2}. The roles are reversed for $k_{||}$$<$0.  Figure~S2
~of SM~\cite{supp} shows the evolution of the  CDW gap across the BZ; a good agreement of ARPES and EBS is also apparent here,  see Discussion~A
~of SM~\cite{supp}. 

\subsection{Main and Shadow bands towards $k_z$ near the BZ boundary}
\label{subsec:YT_shadow_bands_KNL}

\begin{figure*}[!tb] 
\includegraphics[width=\textwidth,keepaspectratio,trim={0 0 0 0 },clip]{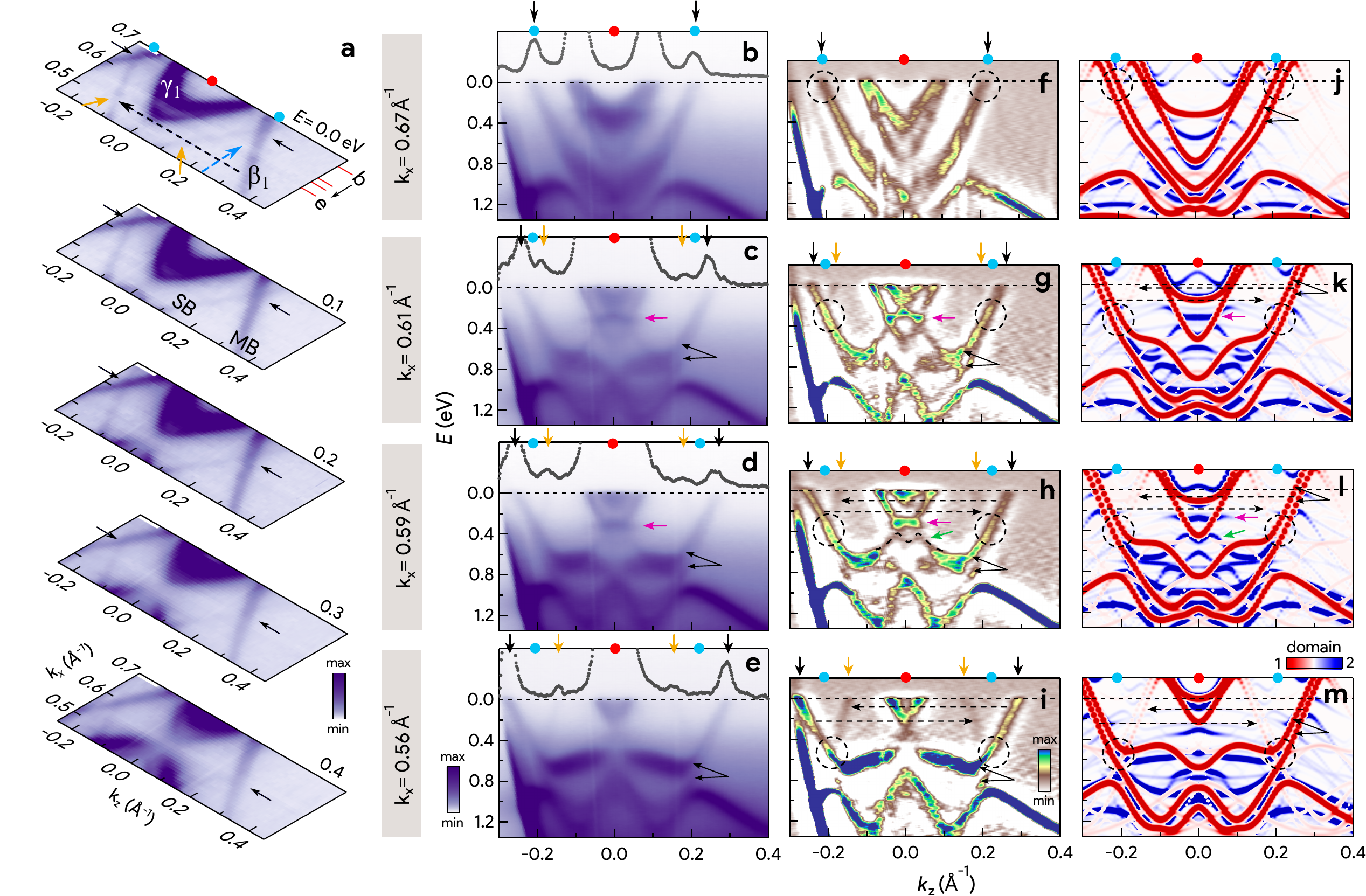}
\caption{
	(a) A series of $k_x$-$k_z$ isosurface plots  with $E$ varying from 0.0 to  0.4 eV measured with h$\nu$= 28 eV around the $X$ point (red dot), showing the crossings (black arrows) between the main and shadow branches. $E(k_z)$ ARPES intensity plots at  
	$k_x$=  (b) 0.67, (c) 0.61, (d) 0.59, and (e) 0.56 \a.  The momentum distribution curves measured at the \ef\ are shown by the gray filled circles above the dashed horizontal line in  panels \textbf{b-e}. The peaks related to the main and  shadow bands are indicated by the black and yellow vertical arrows, respectively. The bilayer splitting is also indicated by a pair of black arrows. 
	~The color scale for panels \textbf{a-e} is shown on the left of panel \textbf{e}. (f-i) 2D curvature plots of panels \textbf{b-e}, respectively. The  crossing regions are highlighted by  black dashed circles. The red and cyan dots at the top of the panels indicate the $\Gamma X$ and $\Gamma_2 X_2$  lines, respectively. The dashed black arrows in panel \textbf{h} denote \q\ connecting the main bands and shadow bands. The pink arrows in panels \textbf{c, d, g, h} indicate flat bands observed around $k_z$= 0. The green arrow in panel \textbf{h} indicates the $``$M$"$ shaped bands, where a dashed curve serves as guide to the eye. 	The color scale is shown within the panel \textbf{i}.   (j-m) Calculated EBS at similar $k_x$ values as panels \textbf{b-e}, where bands from both the domains [\textit{1} (red) and \textit{2} (blue)] are overlaid.  The possible crossing region of the shadow and main bands is highlighted by black dashed circles. The color scale is shown at the top of panel \textbf{m}. }
\label{YTe3_crossing}	
\end{figure*}

In this subsection, we investigate the interaction between the ARPES main bands that are present also in the non-CDW state  and shadow bands  that only appear in the CDW state.   Fig.~\ref{YTe3_crossing}(a) shows a stack of $k_z$-$k_x$ isosurface plots for different $E$  ranging from $E$= 0 to 0.4 eV. 
~The FS i.e., the top plot for $E$= 0 as well as other isosurface plots at larger $E$ show the  shadow branches  corresponding to the main $\beta_1$ branch, as indicated by yellow arrows. The  main and shadow branches appear to cross,  and the crossing point shifts to lower $k_x$ as $E$ \ssA{increases (shown by the  black arrows)}. Notably, the crossing point does not shift in $k_z$ and occurs at $k_z$= 0.209~\a, (blue dashed arrow). Additionally,  the crossings between the main  and shadow bands are also identified in the  ARPES intensity plots, $E$($k_z$),   measured at $k_x$= 0.67, 0.61, 0.59, and 0.56~\a, respectively [Figs.~\ref{YTe3_crossing}(b-e)].  The red horizontal lines on the right axis of the FS  in Fig.~\ref{YTe3_crossing}(a) indicate these directions.  In Fig.~\ref{YTe3_crossing}(b), we observe two main  bands (inner and outer) of parabolic shape that are centered around the $X$ point (indicated by the red dot at the top horizontal axis). These bands intersect  \ef\ at  $k_z$$\approx$$\pm$0.15 and $\pm$0.21~\a. The outer main band related to the $\beta_1$ main branch disperses down to an energy of $\sim$1.2 eV, while the inner main band related to the $\gamma_1$ main branch has the minimum at $\sim$0.8 eV.   With decreasing $k_x$ in Figs.~\ref{YTe3_crossing}(c-e), the  outer main band spreads out in $k_z$, while the inner bands shrink in $k_z$ and its bottom moves towards lower $E$.  Additionally, a flat band at $E$= 0.3 eV [\ssA{pink} 
~arrows in Figs.~\ref{YTe3_crossing}(c,d) and (g,h)] along with a $``$M$"$-shaped band centered around $k_z$= 0 [green arrow in Fig.~\ref{YTe3_crossing}(h), dashed black curve is guide to the eye] partly overlap with the main bands.  

From the EBS  in Figs.~\ref{YTe3_crossing}(j-m)  [calculated at same $k_x$ values as in Figs.~\ref{YTe3_crossing}(f-i), respectively], where bands of both the domains are overlaid, we find that these additional  bands (blue color) are related to  domain~\textit{2} [pink and green arrows in Fig.~\ref{YTe3_crossing}(l)]. Furthermore, EBS reveals a splitting in main band, as highlighted by double black arrows in Fig.~\ref{YTe3_crossing}(k). This  splitting -- referred to as  ``bilayer splitting"-- occurs due to the interaction between the Te bilayers that host the CDW in \yt\ and has been observed in other RTe$_3$ members~\cite{Sarkar2023,Brouet2008, Gweon1998,Chikina2023}. Additionally, the calculated band structure $E(k_z)$ at $k_x$= 0.59 \a (gray curves in Fig.~S3
~of SM~\cite{supp}) illustrates that the bilayer splitting for the outer main band increases with energy. For example, at $E$= 0 eV, the splitting ($\Delta k_z$) is 0.2 \a, while at $E$= 0.4 eV, it is 0.3 \a. Note that bilayer splitting is not related to the CDW, as has been discussed in detail for \lt~\cite{Sarkar2023}. For \yt\ also,  a comparison of the CDW band structures with the non-CDW state (blue dashed) in Fig.~S3
~of SM~\cite{supp} reveals that the bilayer splitting is comparable in both states.  

The ARPES intensity plots are however complicated by the  presence of twin domains, bands from domain \textit{2}  interfere with the bands from domain~\textit{1}. For example,  some of the blue bands from domain \textit{2} can be seen in between the bilayer split main bands and overlapping with the crossing region within the black dashed circles in Figs.~\ref{YTe3_crossing}(k-m). Nevertheless, bilayer splitting is observed at higher $E$, where it is pronounced [two black arrows in Figs.~\ref{YTe3_crossing}(c-e) and (g-i)]. 

\begin{figure*}[!ht] 
	\includegraphics[width=\columnwidth,keepaspectratio,trim={0 0 0 0 },clip]{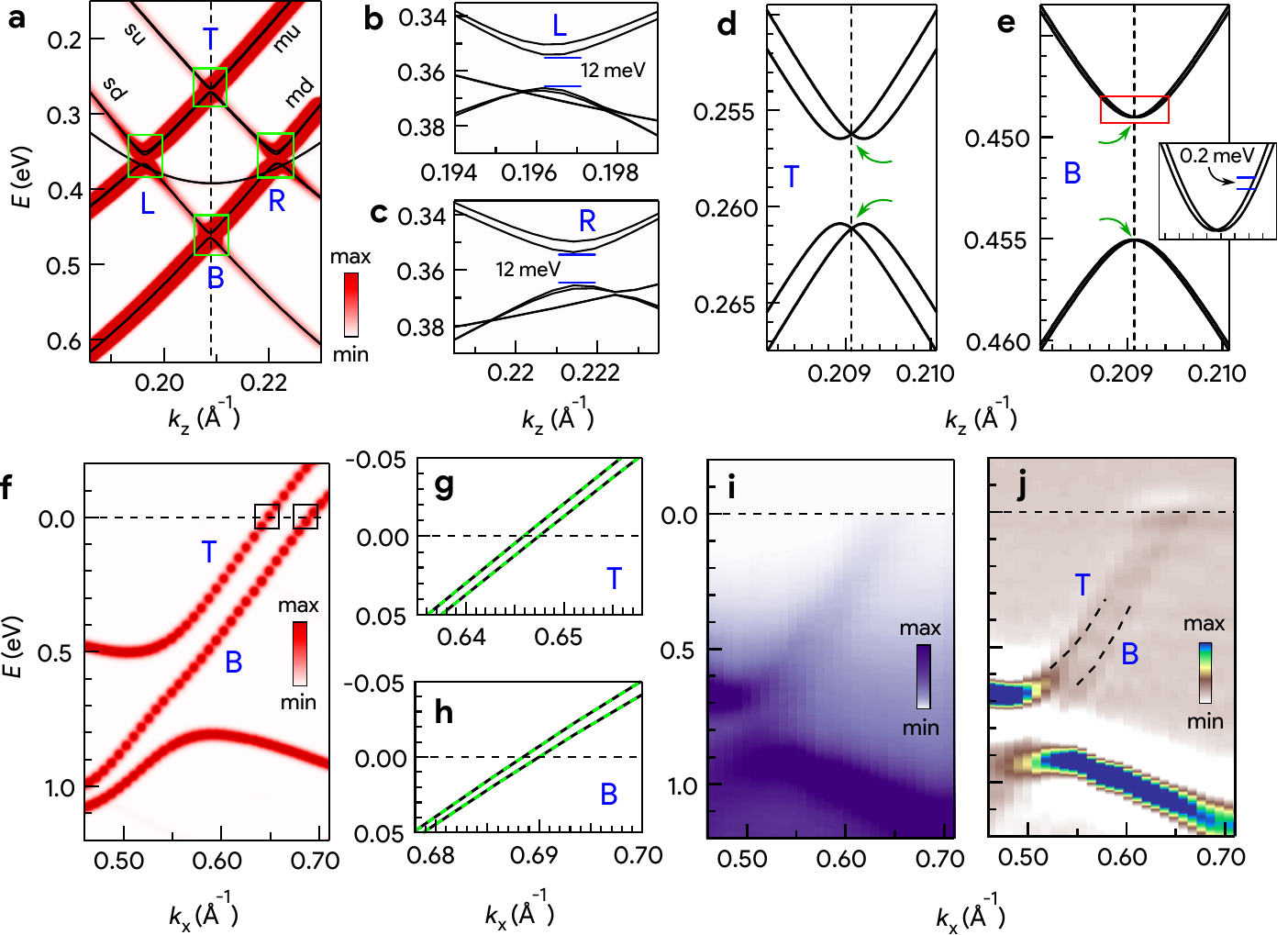}
	\caption{ (a) $E$(k$_z$) bands with SOC at k$_x$= 0.59 \a\ in the crossing region of the main and shadow bands. The vertical dashed line represents the k$_z$ point	on $\Gamma_2$X$_2$ i.e., the $\Sigma$ line in the second BZ [see Fig.~\ref{YTe3_FS}(a)]. \ssA{The black lines  represent the folded bands, whereas the EBS is shown in red color scale}. Expanded regions around (b) \textit{L}, (c) \textit{R}, (d) \textit{T}, and (e) \textit{B} crossings. The inset in panel \textbf{e} shows the expanded region inside the red rectangle and the green arrows in panels \textbf{d,e} indicate the doubly degenerate crossings. (f) EBS along the $\Gamma_2 \rm{X}_2$ line showing the \textit{T} and \textit{B} related bands traversing the Fermi level. The band $E(k_x)$ dispersion from DFT shown in an expanded scale  for the (g) \textit{T} and (h) \textit{B} related bands. (i) The corresponding $E(k_x$) ARPES intensity plot  towards $\Gamma_2 \rm{X}_2$  [i.e., along the blue dashed arrow in the top plot ($E$= 0) in Fig.~\ref{YTe3_crossing}(a)].  (j) 2D curvature plot of panel \textbf{i}.} 
	\label{YTe3_LRTB}	
\end{figure*}

The evidence of  shadow bands is obtained in the 2D curvature plots  e.g., in Figs.~\ref{YTe3_crossing}(g-i). The shadow bands  on both sides of  the \textit{X} point are separated by \q\ from the corresponding  main bands, as shown by the  black  dashed horizontal arrows.  In the  raw APRES intensity plots in Figs.~\ref{YTe3_crossing}(b-e), the momentum distribution curves (MDC) at  the \ef\  (gray filled circles)  depicts the $k_z$ where the shadow bands   cross \ef\ (yellow vertical arrows). The larger intensity peaks in the MDC are related to the main bands (black vertical arrows). As $k_x$ decreases, the $k_z$ separation between the main  and the shadow band increases, and their  crossing -- indicated by the black dashed circles -- shifts to higher $E$ values. However, independent of $k_x$, the crossings occur at same $k_z$  at the $\Gamma_2 X_2$ line [cyan dot at the top of Figs.~\ref{YTe3_crossing}(f-i)].

It may be noted that, in good agreement with ARPES, the calculated EBSs in Figs.~\ref{YTe3_crossing}(j-m) show that the shadow bands  are  separated by \q\ from the main bands, as  indicated by the horizontal black dashed arrows.   EBS clearly shows that the shadow bands in \yt\  are weaker than in \lt~\cite{Sarkar2023}, which is consistent with an observation from ARPES. So, this is not related to extraneous factors, but  is related to the difference in the crystal structure of these two compounds. The most notable difference  -- although their \qcdw\ are close -- is in the amplitude ($A$) of the CDW modulation that is a factor of two smaller in \yt\ ($A$= 0.07~\AA\ from Fig.~S1
~of SM~\cite{supp}) compared to \lt\ ($A$= 0.14~\AA\ from the supplementary Fig.~2 of Ref.~\onlinecite{Sarkar2023}). Additionally, our  EBS calculation for \lt\ showed that if  $A$ decreases  to about one-fifth of the experimental value, the shadow bands are almost completely absent~\cite{Sarkar2023}. Thus, it is reasonable to conclude that the weakness of the shadow bands in \yt\ is related to the smaller amplitude of  the CDW.  In spite of this, the  crossings of the shadow bands with  the main bands  
~are observed [highlighted by the black dashed circles in Figs. \ref{YTe3_crossing}(j-m)]. These show similar trend as ARPES, i.e., the crossings move towards larger $E$ with lower $k_x$. 

\subsection{Evidence of Kramers nodal line in \yt}

Figure~\ref{YTe3_LRTB}(a)  shows the EBS 
~of the crossing region at $k_x$= 0.59 \a\  corresponding to the black dashed circle in Fig.~\ref{YTe3_crossing}(l). EBS shows four  crossings (\textit{L}, \textit{R}, \textit{T} and \textit{B}) between the bilayer split main band (\textit{mu} and \textit{md}) and the shadow band (\textit{su} and \textit{sd}). Since the electronic structure of the structurally similar domains would be same, 
~to avoid the interference from the bands of domain \textit{2}, we have  shown the EBS for domain \textit{1} only and the corresponding folded band structure (black curves) is overlaid on it. Note that the additional parabolic band due to band folding is not detected in the unfolded EBS because of its reduced spectral weight and is also not observed in ARPES [Figs.~\ref{YTe3_crossing}(d,h)]. Notably, although all the four crossings were observed from ARPES in  \lt~\cite{Sarkar2023}, in case of \yt\ the weakness of shadow bands  and interference of the twin domains make it difficult to  identify them.

In Figs.~\ref{YTe3_LRTB}(b,c), expanded regions near \textit{L} and \textit{R} crossings of \yt\ show that both  are gapped by 12 meV.  $T$ and $B$ also exhibit a mini gap  (4-6 meV).  To decipher the influence of SOC, we have  calculated the folded band structure near the crossing region without SOC (Fig.~S4
~of SM~\cite{supp}). Mini-gaps are observed at  $T$ and $B$  without SOC, signifying the hybridization of  bands involved in these crossings. In contrast, $L$ and $R$ exhibit gapless crossings without SOC. Thus nodal lines are formed at $L$ and $R$ that are gapped out with SOC. A similar behavior has been reported for  \lt~\cite{Sarkar2023}. \ssA{When SOC is considered, each of the gapped branches of both $T$ and $B$  exhibit  splitting that is expected in a  noncentrosymmetric system such as \yt.} 
However, it is interesting to note  that the bands cross at $k_z$= 0.209 \a, as indicated by green arrows in Figs.~\ref{YTe3_LRTB}(d,e).  $k_z$= 0.209~\a\ is significant because it corresponds to the $\Gamma_2 X_2$ direction. The origin of the crossings is discussed later.
~The SOC splitting is more pronounced in case of bands at the $T$ crossing [2 meV, Fig.~\ref{YTe3_LRTB}(d)] compared to the $B$ crossing [0.2 meV, inset of Fig.~\ref{YTe3_LRTB}(e)]. 

\begin{figure}[!tb] 
	\includegraphics[width=\columnwidth,keepaspectratio,trim={0 0 0 0 },clip]{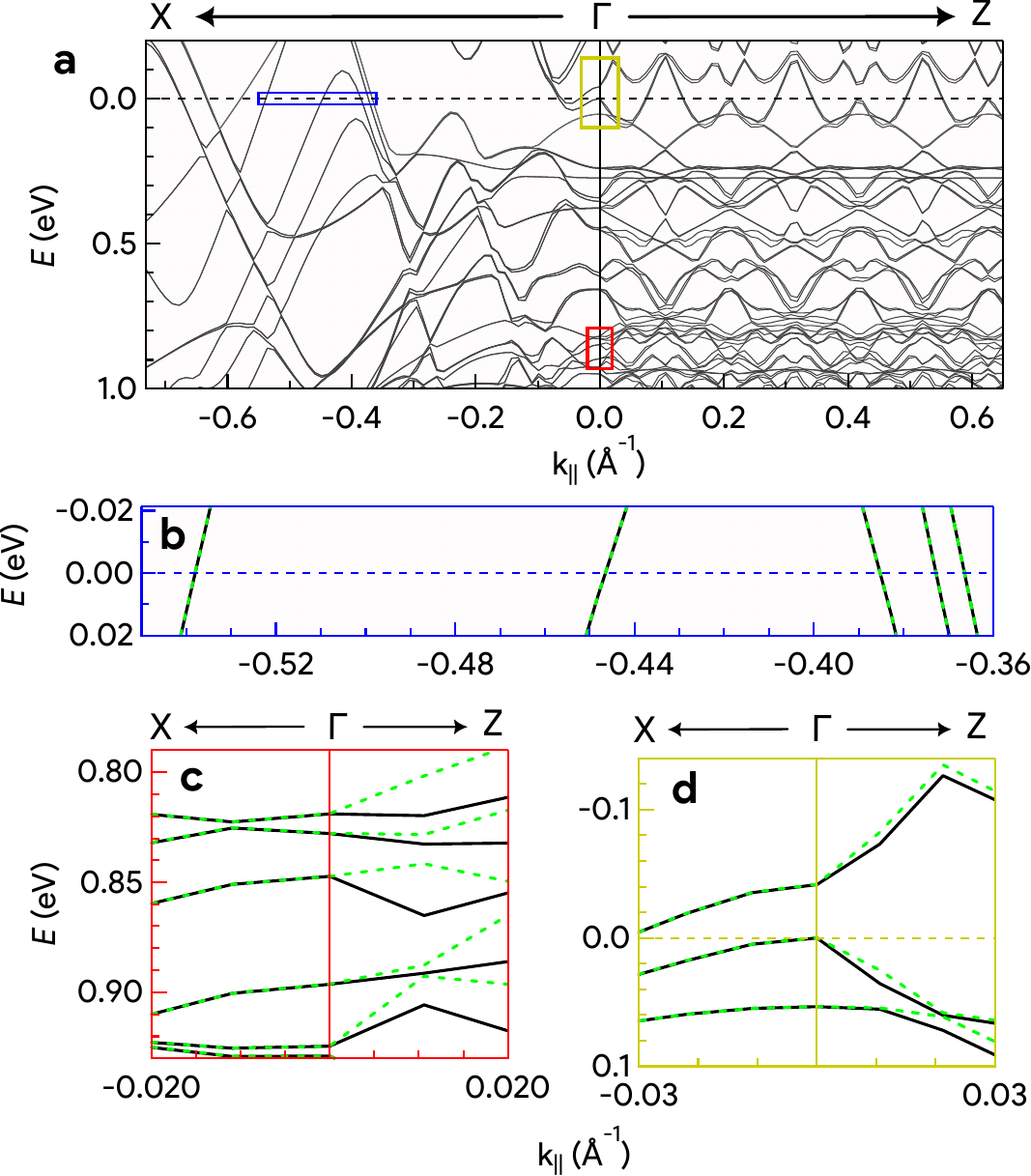}
\caption{ (a) DFT calculated folded band structure of seven-fold \yt\  along $X$$\Gamma$$Z$ with SOC. (b) A zoomed region from the blue rectangle in panel \textbf{a} shows the two fold degeneracy (dashed-green and black) of the bands along $\Gamma$$X$ i.e., the KNL. Zoomed regions from (c) the red and (d)  yellow rectangles in panel \textbf{a} show the spin-splitting along the $\Gamma Z$ direction, where as all bands along the KNL are degenerate.} 
\label{YTe3_XGZ_degeneracy}	
\end{figure}

 In Fig.~\ref{YTe3_LRTB}(f), the EBS  along $\Gamma_2$X$_2$ direction shows the dispersion of  \textit{T} and \textit{B} crossing related bands, i.e., these bands trace the crossings along the $k_x$ direction. However, due to the  broadening used, we do not observe both the branches. Both the mini-gapped branches   of  \textit{T} and \textit{B} are however evident in the folded band structure shown zoomed in Figs.~\ref{YTe3_LRTB}(g) and (h), respectively, where  two doubly degenerate branches cross the \ef.   From the $E(k_x)$ ARPES intensity plot and its 2D curvature plot in Figs.~\ref{YTe3_LRTB}(i,j) towards the $\Gamma_2 X_2$ direction at $k_z$= 0.209~\a, 
 ~the \textit{T} and \textit{B} crossing related bands are  experimentally observed. These cross \ef\ at $k_x$= 0.62 and 0.66~\a, respectively. 
~It is worth mentioning that the EBS  [Fig.~\ref{YTe3_LRTB}(f)] exhibits excellent agreement with the ARPES bands in Figs.~\ref{YTe3_LRTB}(i,j). 

An analysis of the orbital texture of the EBS for $k_x$= 0.59~\a\  \ssA{[Fig.~\ref{YTe3_LRTB}(a)]} in Figs.~S5(a-c)
~of SM~\cite{supp}, shows that both the main and shadow bands are predominantly of Te $p_x$--$p_z$ character near the crossing regions, while Te $p_y$ character becomes significant for E$\geq$0.5 eV, indicating the 2D character of both the main and shadow bands. Further, the orbital textures of the EBS along $\Gamma_2 X_2$ \ssA{[Fig.~\ref{YTe3_LRTB}(f)]} presented in Figs.~S5(d-f)
~of SM~\cite{supp} reveal that both \textit{T} and \textit{B} bands exhibit significant contributions from the in-plane Te orbitals ($p_x$ and $p_z$), showcasing their 2D nature. However, above $E$= 0.4 eV, there is some $p_y$ contribution in the \textit{T} band. 

The double degeneracy of the bands  discussed above is in fact true for all the DFT calculated bands along  $\Gamma X$,   as shown in  Fig.~\ref{YTe3_XGZ_degeneracy}. The zoomed blue rectangular region in  Fig.~\ref{YTe3_XGZ_degeneracy}(b) shows this for the different bands crossing the \ef.  The zoomed yellow  and red rectangle regions in Figs.~\ref{YTe3_XGZ_degeneracy}(c,d) show, in contrast,  that the degeneracy is lifted by the SOC along other directions $i.e.$, along $\Gamma Z$,  whereas it remains in tact along $\Gamma X$ both near \ef\ and at larger $E$, respectively.  The seven-fold structure of \yt\ with \textit{C2cm} space group and the little group that is isomorphic to $C_{2v}$ point group, coupled with protection from time reversal symmetry, enforces the double degeneracy along $\Gamma_2 X_2$, $i.e.$  $\Gamma X$ or the $\Sigma$ line~\cite{Sarkar2023}. This is the signature of the Kramers nodal line (KNL) \cite{Xie2021} that always connects two TRIMs.  In the present case,  we find that both the upper and lower branches of $T$ and $B$ exhibit quadratic dispersion in the vicinity of the crossings [Figs.~\ref{YTe3_LRTB}(d,e)]. This type of crossing with quadratic and cubic dispersion  has been described as higher-order Dirac points (two fold degenerate) and  the Berry phase around a KNL is quantized as m$\pi$ mod 2$\pi$~\cite{Xie2021}. The crossings of the CDW-induced shadow with  the main bands  are enforced by the KNL. The crossings of $E(k_z)$ bands disperse and traverse  the \ef\, as $k_x$ increases [Figs.~\ref{YTe3_LRTB}(f-j)] and multiple degenerate $E(k_x)$ bands along the KNL cross \ef\ [Figs.~\ref{YTe3_XGZ_degeneracy}(a,b)]  establishing that \yt\ is a KNL metal.

\section{Conclusions}
We present a comprehensive investigation of a non-magnetic chalcogenide  material \yt\ in the CDW state combining results of the electronic band structure using ARPES, DFT performed with an experiment-based structure determined by x-ray crystallography and surface properties  using STM and LEED.  We find that the crystal structure of  \yt\  is noncentrosymmetric with $C2cm$ basic space group and  \qcdw\ of 0.2907c$^*$. A commensurate fiftyfive-fold structure is derived from the atomic positions determined by x-ray crystallography, which has a \qcdw\ indistinguishable from the x-ray crystallography value within the experimental accuracy. 
~However, since this unit cell is very large, the DFT calculations have been performed with a seven-fold structure, albeit with same symmetries. 

Evidence of two mutually orthogonal CDW domains, each exhibiting an unidirectional and similar  \qcdw\ is obtained from  both LEED and STM.   The EBS, obtained by unfolding the DFT calculated bands, provides excellent agreement with ARPES when the contributions from both the  twin domains are considered. The CDW gap is barely visible in the FS of \yt\ because of  the twin domains. Nevertheless, the gap could be identified by comparing with the FS measured at high temperature in the non-CDW state. Thus, twin domains influence the band structure  and the Fermi surface of \yt. 

 CDW modulation related weak shadow bands near the BZ boundary, which are shifted from the main bands by \qcdw, have been identified in ARPES intensity plots and are further corroborated by the  EBS calculation.  Additionally, possible existence of multiple crossings between the bilayer split main and shadow bands is portrayed both by experiment and theory. The bands calculated using DFT with SOC show a nodal line along $\Sigma$ resulting from the crossings between the bilayer split main and shadow bands perpendicular to the KNL. Doubly degenerate bands are found only along the KNL for all energy values, with certain bands crossing the \ef.  The KNL  exists only in the CDW phase where shadow bands are present and the inversion symmetry is broken. The symmetries of the \textit{C2cm} space group and the time reversal symmetry  protect the KNL.   Thus, our present work  establishes that \yt\ is a  KNL metal.  The intricate interplay between CDW,  non-trivial topology and twin domains  in \yt\ unveils a rich landscape for further exploration and offers a pathway to deeper insights into exotic topological phases in condensed matter systems. 
\section{Acknowledgments}
S.S., P.S. and S.R.B.  gratefully acknowledge the  financial support from  Department of Science and Technology, Government of India within  the framework of the DST-Synchrotron-Neutron Project to perform experiments at ASTRID2 synchrotron facility.    A part of this work was supported by VILLUM FONDEN via the Centre of Excellence for Dirac Materials (Grant No. 11744). H.~S. Kunwar is thanked for help in the Raman measurement. The Computer division of Raja Ramanna Centre for Advanced Technology  is thanked for installing the DFT codes and providing support throughout.  At Argonne, this work was supported by the U.S. Department of Energy, Basic Energy Sciences, Office of Science, Materials Sciences and Engineering Division (synthesis, crystals growth, property characterization).



\noindent $^{*}$shuvamsarkarhere@gmail.com\\
{$^{\dagger}$barmansr@gmail.com}


\begin{thebibliography}{45}%
	\makeatletter
	\providecommand \@ifxundefined [1]{%
		\@ifx{#1\undefined}
	}%
	\providecommand \@ifnum [1]{%
		\ifnum #1\expandafter \@firstoftwo
		\else \expandafter \@secondoftwo
		\fi
	}%
	\providecommand \@ifx [1]{%
		\ifx #1\expandafter \@firstoftwo
		\else \expandafter \@secondoftwo
		\fi
	}%
	\providecommand \natexlab [1]{#1}%
	\providecommand \enquote  [1]{``#1''}%
	\providecommand \bibnamefont  [1]{#1}%
	\providecommand \bibfnamefont [1]{#1}%
	\providecommand \citenamefont [1]{#1}%
	\providecommand \href@noop [0]{\@secondoftwo}%
	\providecommand \href [0]{\begingroup \@sanitize@url \@href}%
	\providecommand \@href[1]{\@@startlink{#1}\@@href}%
	\providecommand \@@href[1]{\endgroup#1\@@endlink}%
	\providecommand \@sanitize@url [0]{\catcode `\\12\catcode `\$12\catcode
		`\&12\catcode `\#12\catcode `\^12\catcode `\_12\catcode `\%12\relax}%
	\providecommand \@@startlink[1]{}%
	\providecommand \@@endlink[0]{}%
	\providecommand \url  [0]{\begingroup\@sanitize@url \@url }%
	\providecommand \@url [1]{\endgroup\@href {#1}{\urlprefix }}%
	\providecommand \urlprefix  [0]{URL }%
	\providecommand \Eprint [0]{\href }%
	\providecommand \doibase [0]{https://doi.org/}%
	\providecommand \selectlanguage [0]{\@gobble}%
	\providecommand \bibinfo  [0]{\@secondoftwo}%
	\providecommand \bibfield  [0]{\@secondoftwo}%
	\providecommand \translation [1]{[#1]}%
	\providecommand \BibitemOpen [0]{}%
	\providecommand \bibitemStop [0]{}%
	\providecommand \bibitemNoStop [0]{.\EOS\space}%
	\providecommand \EOS [0]{\spacefactor3000\relax}%
	\providecommand \BibitemShut  [1]{\csname bibitem#1\endcsname}%
	\let\auto@bib@innerbib\@empty
	\bibitem [{\citenamefont {Sarkar}\ \emph {et~al.}(2023)\citenamefont {Sarkar},
		\citenamefont {Bhattacharya}, \citenamefont {Sadhukhan}, \citenamefont
		{Curcio}, \citenamefont {Dutt}, \citenamefont {Singh}, \citenamefont
		{Bianchi}, \citenamefont {Pariari}, \citenamefont {Roy}, \citenamefont
		{Mandal}, \citenamefont {Das}, \citenamefont {Hofmann}, \citenamefont
		{Chakrabarti},\ and\ \citenamefont {Barman}}]{Sarkar2023}%
	\BibitemOpen
	\bibfield  {author} {\bibinfo {author} {\bibfnamefont {S.}~\bibnamefont
			{Sarkar}}, \bibinfo {author} {\bibfnamefont {J.}~\bibnamefont
			{Bhattacharya}}, \bibinfo {author} {\bibfnamefont {P.}~\bibnamefont
			{Sadhukhan}}, \bibinfo {author} {\bibfnamefont {D.}~\bibnamefont {Curcio}},
		\bibinfo {author} {\bibfnamefont {R.}~\bibnamefont {Dutt}}, \bibinfo {author}
		{\bibfnamefont {V.~K.}\ \bibnamefont {Singh}}, \bibinfo {author}
		{\bibfnamefont {M.}~\bibnamefont {Bianchi}}, \bibinfo {author} {\bibfnamefont
			{A.}~\bibnamefont {Pariari}}, \bibinfo {author} {\bibfnamefont
			{S.}~\bibnamefont {Roy}}, \bibinfo {author} {\bibfnamefont {P.}~\bibnamefont
			{Mandal}}, \bibinfo {author} {\bibfnamefont {T.}~\bibnamefont {Das}},
		\bibinfo {author} {\bibfnamefont {P.}~\bibnamefont {Hofmann}}, \bibinfo
		{author} {\bibfnamefont {A.}~\bibnamefont {Chakrabarti}},\ and\ \bibinfo
		{author} {\bibfnamefont {S.~R.}\ \bibnamefont {Barman}},\ }\bibfield  {title}
	{\bibinfo {title} {Charge density wave induced nodal lines in {LaTe}$_3$},\
	}\href {https://doi.org/10.1038/s41467-023-39271-1} {\bibfield  {journal}
		{\bibinfo  {journal} {Nat. Commun.}\ }\textbf {\bibinfo {volume} {14}},\
		\bibinfo {pages} {3628} (\bibinfo {year} {2023})}\BibitemShut {NoStop}%
	\bibitem [{\citenamefont {Li}\ \emph {et~al.}(2021)\citenamefont {Li},
		\citenamefont {Zhang}, \citenamefont {Yilmaz}, \citenamefont {Pai},
		\citenamefont {Marvinney}, \citenamefont {Said}, \citenamefont {Yin},
		\citenamefont {Gong}, \citenamefont {Tu}, \citenamefont {Vescovo},
		\citenamefont {Nelson}, \citenamefont {Moore}, \citenamefont {Murakami},
		\citenamefont {Lei}, \citenamefont {Lee}, \citenamefont {Lawrie},\ and\
		\citenamefont {Miao}}]{Li2021}%
	\BibitemOpen
	\bibfield  {author} {\bibinfo {author} {\bibfnamefont {H.}~\bibnamefont
			{Li}}, \bibinfo {author} {\bibfnamefont {T.~T.}\ \bibnamefont {Zhang}},
		\bibinfo {author} {\bibfnamefont {T.}~\bibnamefont {Yilmaz}}, \bibinfo
		{author} {\bibfnamefont {Y.~Y.}\ \bibnamefont {Pai}}, \bibinfo {author}
		{\bibfnamefont {C.~E.}\ \bibnamefont {Marvinney}}, \bibinfo {author}
		{\bibfnamefont {A.}~\bibnamefont {Said}}, \bibinfo {author} {\bibfnamefont
			{Q.~W.}\ \bibnamefont {Yin}}, \bibinfo {author} {\bibfnamefont {C.~S.}\
			\bibnamefont {Gong}}, \bibinfo {author} {\bibfnamefont {Z.~J.}\ \bibnamefont
			{Tu}}, \bibinfo {author} {\bibfnamefont {E.}~\bibnamefont {Vescovo}},
		\bibinfo {author} {\bibfnamefont {C.~S.}\ \bibnamefont {Nelson}}, \bibinfo
		{author} {\bibfnamefont {R.~G.}\ \bibnamefont {Moore}}, \bibinfo {author}
		{\bibfnamefont {S.}~\bibnamefont {Murakami}}, \bibinfo {author}
		{\bibfnamefont {H.~C.}\ \bibnamefont {Lei}}, \bibinfo {author} {\bibfnamefont
			{H.~N.}\ \bibnamefont {Lee}}, \bibinfo {author} {\bibfnamefont {B.~J.}\
			\bibnamefont {Lawrie}},\ and\ \bibinfo {author} {\bibfnamefont
			{H.}~\bibnamefont {Miao}},\ }\bibfield  {title} {\bibinfo {title}
		{Observation of unconventional charge density wave without acoustic phonon
			anomaly in kagome superconductors {AV}$_3${S}b$_5$ (${A}=\mathrm{Rb}$,
			{Cs})},\ }\href {https://doi.org/10.1103/PhysRevX.11.031050} {\bibfield
		{journal} {\bibinfo  {journal} {Phys. Rev. X}\ }\textbf {\bibinfo {volume}
			{11}},\ \bibinfo {pages} {031050} (\bibinfo {year} {2021})}\BibitemShut
	{NoStop}%
	\bibitem [{\citenamefont {Shi}\ \emph {et~al.}(2021)\citenamefont {Shi},
		\citenamefont {Wieder}, \citenamefont {Meyerheim}, \citenamefont {Sun},
		\citenamefont {Zhang}, \citenamefont {Li}, \citenamefont {Shen},
		\citenamefont {Qi}, \citenamefont {Yang}, \citenamefont {Jena}, \citenamefont
		{Werner}, \citenamefont {Koepernik}, \citenamefont {Parkin}, \citenamefont
		{Chen}, \citenamefont {Felser}, \citenamefont {Bernevig},\ and\ \citenamefont
		{Wang}}]{Shi2021}%
	\BibitemOpen
	\bibfield  {author} {\bibinfo {author} {\bibfnamefont {W.}~\bibnamefont
			{Shi}}, \bibinfo {author} {\bibfnamefont {B.~J.}\ \bibnamefont {Wieder}},
		\bibinfo {author} {\bibfnamefont {H.~L.}\ \bibnamefont {Meyerheim}}, \bibinfo
		{author} {\bibfnamefont {Y.}~\bibnamefont {Sun}}, \bibinfo {author}
		{\bibfnamefont {Y.}~\bibnamefont {Zhang}}, \bibinfo {author} {\bibfnamefont
			{Y.}~\bibnamefont {Li}}, \bibinfo {author} {\bibfnamefont {L.}~\bibnamefont
			{Shen}}, \bibinfo {author} {\bibfnamefont {Y.}~\bibnamefont {Qi}}, \bibinfo
		{author} {\bibfnamefont {L.}~\bibnamefont {Yang}}, \bibinfo {author}
		{\bibfnamefont {J.}~\bibnamefont {Jena}}, \bibinfo {author} {\bibfnamefont
			{P.}~\bibnamefont {Werner}}, \bibinfo {author} {\bibfnamefont
			{K.}~\bibnamefont {Koepernik}}, \bibinfo {author} {\bibfnamefont
			{S.}~\bibnamefont {Parkin}}, \bibinfo {author} {\bibfnamefont
			{Y.}~\bibnamefont {Chen}}, \bibinfo {author} {\bibfnamefont {C.}~\bibnamefont
			{Felser}}, \bibinfo {author} {\bibfnamefont {B.~A.}\ \bibnamefont
			{Bernevig}},\ and\ \bibinfo {author} {\bibfnamefont {Z.}~\bibnamefont
			{Wang}},\ }\bibfield  {title} {\bibinfo {title} {A charge-density-wave
			topological semimetal},\ }\href {https://doi.org/10.1038/s41567-020-01104-z}
	{\bibfield  {journal} {\bibinfo  {journal} {Nat. Phys.}\ }\textbf {\bibinfo
			{volume} {17}},\ \bibinfo {pages} {381} (\bibinfo {year} {2021})}\BibitemShut
	{NoStop}%
	\bibitem [{\citenamefont {Qian}\ \emph {et~al.}(2014)\citenamefont {Qian},
		\citenamefont {Liu}, \citenamefont {Fu},\ and\ \citenamefont
		{Li}}]{Qian2014}%
	\BibitemOpen
	\bibfield  {author} {\bibinfo {author} {\bibfnamefont {X.}~\bibnamefont
			{Qian}}, \bibinfo {author} {\bibfnamefont {J.}~\bibnamefont {Liu}}, \bibinfo
		{author} {\bibfnamefont {L.}~\bibnamefont {Fu}},\ and\ \bibinfo {author}
		{\bibfnamefont {J.}~\bibnamefont {Li}},\ }\bibfield  {title} {\bibinfo
		{title} {Quantum spin hall effect in two-dimensional transition metal
			dichalcogenides},\ }\href {https://doi.org/10.1126/science.1256815}
	{\bibfield  {journal} {\bibinfo  {journal} {Science}\ }\textbf {\bibinfo
			{volume} {346}},\ \bibinfo {pages} {1344} (\bibinfo {year}
		{2014})}\BibitemShut {NoStop}%
	\bibitem [{\citenamefont {Lei}\ \emph {et~al.}(2021)\citenamefont {Lei},
		\citenamefont {Teicher}, \citenamefont {Topp}, \citenamefont {Cai},
		\citenamefont {Lin}, \citenamefont {Cheng}, \citenamefont {Salters},
		\citenamefont {Rodolakis}, \citenamefont {McChesney}, \citenamefont
		{Lapidus}, \citenamefont {Yao}, \citenamefont {Krivenkov}, \citenamefont
		{Marchenko}, \citenamefont {Varykhalov}, \citenamefont {Ast}, \citenamefont
		{Car}, \citenamefont {Cano}, \citenamefont {Vergniory}, \citenamefont {Ong},\
		and\ \citenamefont {Schoop}}]{Lei2021}%
	\BibitemOpen
	\bibfield  {author} {\bibinfo {author} {\bibfnamefont {S.}~\bibnamefont
			{Lei}}, \bibinfo {author} {\bibfnamefont {S.~M.~L.}\ \bibnamefont {Teicher}},
		\bibinfo {author} {\bibfnamefont {A.}~\bibnamefont {Topp}}, \bibinfo {author}
		{\bibfnamefont {K.}~\bibnamefont {Cai}}, \bibinfo {author} {\bibfnamefont
			{J.}~\bibnamefont {Lin}}, \bibinfo {author} {\bibfnamefont {G.}~\bibnamefont
			{Cheng}}, \bibinfo {author} {\bibfnamefont {T.~H.}\ \bibnamefont {Salters}},
		\bibinfo {author} {\bibfnamefont {F.}~\bibnamefont {Rodolakis}}, \bibinfo
		{author} {\bibfnamefont {J.~L.}\ \bibnamefont {McChesney}}, \bibinfo {author}
		{\bibfnamefont {S.}~\bibnamefont {Lapidus}}, \bibinfo {author} {\bibfnamefont
			{N.}~\bibnamefont {Yao}}, \bibinfo {author} {\bibfnamefont {M.}~\bibnamefont
			{Krivenkov}}, \bibinfo {author} {\bibfnamefont {D.}~\bibnamefont
			{Marchenko}}, \bibinfo {author} {\bibfnamefont {A.}~\bibnamefont
			{Varykhalov}}, \bibinfo {author} {\bibfnamefont {C.~R.}\ \bibnamefont {Ast}},
		\bibinfo {author} {\bibfnamefont {R.}~\bibnamefont {Car}}, \bibinfo {author}
		{\bibfnamefont {J.}~\bibnamefont {Cano}}, \bibinfo {author} {\bibfnamefont
			{M.~G.}\ \bibnamefont {Vergniory}}, \bibinfo {author} {\bibfnamefont {N.~P.}\
			\bibnamefont {Ong}},\ and\ \bibinfo {author} {\bibfnamefont {L.~M.}\
			\bibnamefont {Schoop}},\ }\bibfield  {title} {\bibinfo {title} {Band
			engineering of dirac semimetals using charge density waves},\ }\href
	{https://doi.org/10.1002/adma.202101591} {\bibfield  {journal} {\bibinfo
			{journal} {Adv. Mater.}\ }\textbf {\bibinfo {volume} {33}},\ \bibinfo {pages}
		{2101591} (\bibinfo {year} {2021})}\BibitemShut {NoStop}%
	\bibitem [{\citenamefont {Yang}\ and\ \citenamefont {Kee}(2010)}]{Yang2010}%
	\BibitemOpen
	\bibfield  {author} {\bibinfo {author} {\bibfnamefont {B.-J.}\ \bibnamefont
			{Yang}}\ and\ \bibinfo {author} {\bibfnamefont {H.-Y.}\ \bibnamefont {Kee}},\
	}\bibfield  {title} {\bibinfo {title} {Searching for topological density-wave
			insulators in multiorbital square-lattice systems},\ }\href
	{https://doi.org/10.1103/PhysRevB.82.195126} {\bibfield  {journal} {\bibinfo
			{journal} {Phys. Rev. B}\ }\textbf {\bibinfo {volume} {82}},\ \bibinfo
		{pages} {195126} (\bibinfo {year} {2010})}\BibitemShut {NoStop}%
	\bibitem [{\citenamefont {Huang}\ \emph {et~al.}(2021)\citenamefont {Huang},
		\citenamefont {Xu}, \citenamefont {Singh}, \citenamefont {Hsu}, \citenamefont
		{Hsu}, \citenamefont {Su}, \citenamefont {Bansil},\ and\ \citenamefont
		{Lin}}]{Huang2021}%
	\BibitemOpen
	\bibfield  {author} {\bibinfo {author} {\bibfnamefont {S.-M.}\ \bibnamefont
			{Huang}}, \bibinfo {author} {\bibfnamefont {S.-Y.}\ \bibnamefont {Xu}},
		\bibinfo {author} {\bibfnamefont {B.}~\bibnamefont {Singh}}, \bibinfo
		{author} {\bibfnamefont {M.-C.}\ \bibnamefont {Hsu}}, \bibinfo {author}
		{\bibfnamefont {C.-H.}\ \bibnamefont {Hsu}}, \bibinfo {author} {\bibfnamefont
			{C.}~\bibnamefont {Su}}, \bibinfo {author} {\bibfnamefont {A.}~\bibnamefont
			{Bansil}},\ and\ \bibinfo {author} {\bibfnamefont {H.}~\bibnamefont {Lin}},\
	}\bibfield  {title} {\bibinfo {title} {Aspects of symmetry and topology in
			the charge density wave phase of 1t{\textendash}{TiSe}2},\ }\href
	{https://doi.org/10.1088/1367-2630/ac1bf4} {\bibfield  {journal} {\bibinfo
			{journal} {New J. Phys.}\ }\textbf {\bibinfo {volume} {23}},\ \bibinfo
		{pages} {083037} (\bibinfo {year} {2021})}\BibitemShut {NoStop}%
	\bibitem [{\citenamefont {Polshyn}\ \emph {et~al.}(2021)\citenamefont
		{Polshyn}, \citenamefont {Zhang}, \citenamefont {Kumar}, \citenamefont
		{Soejima}, \citenamefont {Ledwith}, \citenamefont {Watanabe}, \citenamefont
		{Taniguchi}, \citenamefont {Vishwanath}, \citenamefont {Zaletel},\ and\
		\citenamefont {Young}}]{Polshyn2022}%
	\BibitemOpen
	\bibfield  {author} {\bibinfo {author} {\bibfnamefont {H.}~\bibnamefont
			{Polshyn}}, \bibinfo {author} {\bibfnamefont {Y.}~\bibnamefont {Zhang}},
		\bibinfo {author} {\bibfnamefont {M.~A.}\ \bibnamefont {Kumar}}, \bibinfo
		{author} {\bibfnamefont {T.}~\bibnamefont {Soejima}}, \bibinfo {author}
		{\bibfnamefont {P.}~\bibnamefont {Ledwith}}, \bibinfo {author} {\bibfnamefont
			{K.}~\bibnamefont {Watanabe}}, \bibinfo {author} {\bibfnamefont
			{T.}~\bibnamefont {Taniguchi}}, \bibinfo {author} {\bibfnamefont
			{A.}~\bibnamefont {Vishwanath}}, \bibinfo {author} {\bibfnamefont {M.~P.}\
			\bibnamefont {Zaletel}},\ and\ \bibinfo {author} {\bibfnamefont {A.~F.}\
			\bibnamefont {Young}},\ }\bibfield  {title} {\bibinfo {title} {Topological
			charge density waves at half-integer filling of a moir{\'{e}} superlattice},\
	}\href {https://doi.org/10.1038/s41567-021-01418-6} {\bibfield  {journal}
		{\bibinfo  {journal} {Nat. Phys.}\ }\textbf {\bibinfo {volume} {18}},\
		\bibinfo {pages} {42} (\bibinfo {year} {2021})}\BibitemShut {NoStop}%
	\bibitem [{\citenamefont {Zhang}\ \emph {et~al.}(2020)\citenamefont {Zhang},
		\citenamefont {Gu}, \citenamefont {Sun}, \citenamefont {Luo}, \citenamefont
		{Liu}, \citenamefont {Chen}, \citenamefont {Shao}, \citenamefont {Zhang},
		\citenamefont {Li}, \citenamefont {Sun}, \citenamefont {Li}, \citenamefont
		{Li}, \citenamefont {Xue}, \citenamefont {Ge}, \citenamefont {Xing},
		\citenamefont {Comin}, \citenamefont {Zhu}, \citenamefont {Gao},
		\citenamefont {Yan}, \citenamefont {Feng}, \citenamefont {Pan},\ and\
		\citenamefont {Wang}}]{Zhang2020prb}%
	\BibitemOpen
	\bibfield  {author} {\bibinfo {author} {\bibfnamefont {X.}~\bibnamefont
			{Zhang}}, \bibinfo {author} {\bibfnamefont {Q.}~\bibnamefont {Gu}}, \bibinfo
		{author} {\bibfnamefont {H.}~\bibnamefont {Sun}}, \bibinfo {author}
		{\bibfnamefont {T.}~\bibnamefont {Luo}}, \bibinfo {author} {\bibfnamefont
			{Y.}~\bibnamefont {Liu}}, \bibinfo {author} {\bibfnamefont {Y.}~\bibnamefont
			{Chen}}, \bibinfo {author} {\bibfnamefont {Z.}~\bibnamefont {Shao}}, \bibinfo
		{author} {\bibfnamefont {Z.}~\bibnamefont {Zhang}}, \bibinfo {author}
		{\bibfnamefont {S.}~\bibnamefont {Li}}, \bibinfo {author} {\bibfnamefont
			{Y.}~\bibnamefont {Sun}}, \bibinfo {author} {\bibfnamefont {Y.}~\bibnamefont
			{Li}}, \bibinfo {author} {\bibfnamefont {X.}~\bibnamefont {Li}}, \bibinfo
		{author} {\bibfnamefont {S.}~\bibnamefont {Xue}}, \bibinfo {author}
		{\bibfnamefont {J.}~\bibnamefont {Ge}}, \bibinfo {author} {\bibfnamefont
			{Y.}~\bibnamefont {Xing}}, \bibinfo {author} {\bibfnamefont {R.}~\bibnamefont
			{Comin}}, \bibinfo {author} {\bibfnamefont {Z.}~\bibnamefont {Zhu}}, \bibinfo
		{author} {\bibfnamefont {P.}~\bibnamefont {Gao}}, \bibinfo {author}
		{\bibfnamefont {B.}~\bibnamefont {Yan}}, \bibinfo {author} {\bibfnamefont
			{J.}~\bibnamefont {Feng}}, \bibinfo {author} {\bibfnamefont {M.}~\bibnamefont
			{Pan}},\ and\ \bibinfo {author} {\bibfnamefont {J.}~\bibnamefont {Wang}},\
	}\bibfield  {title} {\bibinfo {title} {Eightfold fermionic excitation in a
			charge density wave compound},\ }\href
	{https://doi.org/10.1103/PhysRevB.102.035125} {\bibfield  {journal} {\bibinfo
			{journal} {Phys. Rev. B}\ }\textbf {\bibinfo {volume} {102}},\ \bibinfo
		{pages} {035125} (\bibinfo {year} {2020})}\BibitemShut {NoStop}%
	\bibitem [{\citenamefont {Mitsuishi}\ \emph {et~al.}(2020)\citenamefont
		{Mitsuishi}, \citenamefont {Sugita}, \citenamefont {Bahramy}, \citenamefont
		{Kamitani}, \citenamefont {Sonobe}, \citenamefont {Sakano}, \citenamefont
		{Shimojima}, \citenamefont {Takahashi}, \citenamefont {Sakai}, \citenamefont
		{Horiba}, \citenamefont {Kumigashira}, \citenamefont {Taguchi}, \citenamefont
		{Miyamoto}, \citenamefont {Okuda}, \citenamefont {Ishiwata}, \citenamefont
		{Motome},\ and\ \citenamefont {Ishizaka}}]{Mitsuishi2020}%
	\BibitemOpen
	\bibfield  {author} {\bibinfo {author} {\bibfnamefont {N.}~\bibnamefont
			{Mitsuishi}}, \bibinfo {author} {\bibfnamefont {Y.}~\bibnamefont {Sugita}},
		\bibinfo {author} {\bibfnamefont {M.~S.}\ \bibnamefont {Bahramy}}, \bibinfo
		{author} {\bibfnamefont {M.}~\bibnamefont {Kamitani}}, \bibinfo {author}
		{\bibfnamefont {T.}~\bibnamefont {Sonobe}}, \bibinfo {author} {\bibfnamefont
			{M.}~\bibnamefont {Sakano}}, \bibinfo {author} {\bibfnamefont
			{T.}~\bibnamefont {Shimojima}}, \bibinfo {author} {\bibfnamefont
			{H.}~\bibnamefont {Takahashi}}, \bibinfo {author} {\bibfnamefont
			{H.}~\bibnamefont {Sakai}}, \bibinfo {author} {\bibfnamefont
			{K.}~\bibnamefont {Horiba}}, \bibinfo {author} {\bibfnamefont
			{H.}~\bibnamefont {Kumigashira}}, \bibinfo {author} {\bibfnamefont
			{K.}~\bibnamefont {Taguchi}}, \bibinfo {author} {\bibfnamefont
			{K.}~\bibnamefont {Miyamoto}}, \bibinfo {author} {\bibfnamefont
			{T.}~\bibnamefont {Okuda}}, \bibinfo {author} {\bibfnamefont
			{S.}~\bibnamefont {Ishiwata}}, \bibinfo {author} {\bibfnamefont
			{Y.}~\bibnamefont {Motome}},\ and\ \bibinfo {author} {\bibfnamefont
			{K.}~\bibnamefont {Ishizaka}},\ }\bibfield  {title} {\bibinfo {title}
		{Switching of band inversion and topological surface states by charge density
			wave},\ }\href {https://doi.org/10.1038/s41467-020-16290-w} {\bibfield
		{journal} {\bibinfo  {journal} {Nat. Commun.}\ }\textbf {\bibinfo {volume}
			{11}},\ \bibinfo {pages} {2466} (\bibinfo {year} {2020})}\BibitemShut
	{NoStop}%
	\bibitem [{\citenamefont {Xie}\ \emph {et~al.}(2021)\citenamefont {Xie},
		\citenamefont {Gao}, \citenamefont {Xu}, \citenamefont {Zhang}, \citenamefont
		{Hu}, \citenamefont {Gao},\ and\ \citenamefont {Law}}]{Xie2021}%
	\BibitemOpen
	\bibfield  {author} {\bibinfo {author} {\bibfnamefont {Y.-M.}\ \bibnamefont
			{Xie}}, \bibinfo {author} {\bibfnamefont {X.-J.}\ \bibnamefont {Gao}},
		\bibinfo {author} {\bibfnamefont {X.~Y.}\ \bibnamefont {Xu}}, \bibinfo
		{author} {\bibfnamefont {C.-P.}\ \bibnamefont {Zhang}}, \bibinfo {author}
		{\bibfnamefont {J.-X.}\ \bibnamefont {Hu}}, \bibinfo {author} {\bibfnamefont
			{J.~Z.}\ \bibnamefont {Gao}},\ and\ \bibinfo {author} {\bibfnamefont {K.~T.}\
			\bibnamefont {Law}},\ }\bibfield  {title} {\bibinfo {title} {Kramers nodal
			line metals},\ }\href {https://doi.org/10.1038/s41467-021-22903-9} {\bibfield
		{journal} {\bibinfo  {journal} {Nat. Commun.}\ }\textbf {\bibinfo {volume}
			{12}},\ \bibinfo {pages} {3064} (\bibinfo {year} {2021})}\BibitemShut
	{NoStop}%
	\bibitem [{\citenamefont {Shang}\ \emph {et~al.}(2022)\citenamefont {Shang},
		\citenamefont {Zhao}, \citenamefont {Hu}, \citenamefont {Ma}, \citenamefont
		{Gawryluk}, \citenamefont {Zhu}, \citenamefont {Zhang}, \citenamefont {Zhen},
		\citenamefont {Yu}, \citenamefont {Xu}, \citenamefont {Zhan}, \citenamefont
		{Pomjakushina}, \citenamefont {Shi},\ and\ \citenamefont
		{Shiroka}}]{Shang2022}%
	\BibitemOpen
	\bibfield  {author} {\bibinfo {author} {\bibfnamefont {T.}~\bibnamefont
			{Shang}}, \bibinfo {author} {\bibfnamefont {J.}~\bibnamefont {Zhao}},
		\bibinfo {author} {\bibfnamefont {L.-H.}\ \bibnamefont {Hu}}, \bibinfo
		{author} {\bibfnamefont {J.}~\bibnamefont {Ma}}, \bibinfo {author}
		{\bibfnamefont {D.~J.}\ \bibnamefont {Gawryluk}}, \bibinfo {author}
		{\bibfnamefont {X.}~\bibnamefont {Zhu}}, \bibinfo {author} {\bibfnamefont
			{H.}~\bibnamefont {Zhang}}, \bibinfo {author} {\bibfnamefont
			{Z.}~\bibnamefont {Zhen}}, \bibinfo {author} {\bibfnamefont {B.}~\bibnamefont
			{Yu}}, \bibinfo {author} {\bibfnamefont {Y.}~\bibnamefont {Xu}}, \bibinfo
		{author} {\bibfnamefont {Q.}~\bibnamefont {Zhan}}, \bibinfo {author}
		{\bibfnamefont {E.}~\bibnamefont {Pomjakushina}}, \bibinfo {author}
		{\bibfnamefont {M.}~\bibnamefont {Shi}},\ and\ \bibinfo {author}
		{\bibfnamefont {T.}~\bibnamefont {Shiroka}},\ }\bibfield  {title} {\bibinfo
		{title} {Unconventional superconductivity in topological {K}ramers nodal-line
			semimetals},\ }\href {https://doi.org/10.1126/sciadv.abq6589} {\bibfield
		{journal} {\bibinfo  {journal} {Science Advances}\ }\textbf {\bibinfo
			{volume} {8}},\ \bibinfo {pages} {eabq6589} (\bibinfo {year}
		{2022})}\BibitemShut {NoStop}%
	\bibitem [{\citenamefont {Zhang}\ \emph {et~al.}(2023)\citenamefont {Zhang},
		\citenamefont {Gao}, \citenamefont {Gao}, \citenamefont {Lei}, \citenamefont
		{Ni}, \citenamefont {Oh}, \citenamefont {Huang}, \citenamefont {Yue},
		\citenamefont {Zonno}, \citenamefont {Gorovikov}, \citenamefont {Hashimoto},
		\citenamefont {Lu}, \citenamefont {Denlinger}, \citenamefont {Birgeneau},
		\citenamefont {Kono}, \citenamefont {Wu}, \citenamefont {Law}, \citenamefont
		{Morosan},\ and\ \citenamefont {Yi}}]{Zhang2023}%
	\BibitemOpen
	\bibfield  {author} {\bibinfo {author} {\bibfnamefont {Y.}~\bibnamefont
			{Zhang}}, \bibinfo {author} {\bibfnamefont {Y.}~\bibnamefont {Gao}}, \bibinfo
		{author} {\bibfnamefont {X.-J.}\ \bibnamefont {Gao}}, \bibinfo {author}
		{\bibfnamefont {S.}~\bibnamefont {Lei}}, \bibinfo {author} {\bibfnamefont
			{Z.}~\bibnamefont {Ni}}, \bibinfo {author} {\bibfnamefont {J.~S.}\
			\bibnamefont {Oh}}, \bibinfo {author} {\bibfnamefont {J.}~\bibnamefont
			{Huang}}, \bibinfo {author} {\bibfnamefont {Z.}~\bibnamefont {Yue}}, \bibinfo
		{author} {\bibfnamefont {M.}~\bibnamefont {Zonno}}, \bibinfo {author}
		{\bibfnamefont {S.}~\bibnamefont {Gorovikov}}, \bibinfo {author}
		{\bibfnamefont {M.}~\bibnamefont {Hashimoto}}, \bibinfo {author}
		{\bibfnamefont {D.}~\bibnamefont {Lu}}, \bibinfo {author} {\bibfnamefont
			{J.~D.}\ \bibnamefont {Denlinger}}, \bibinfo {author} {\bibfnamefont {R.~J.}\
			\bibnamefont {Birgeneau}}, \bibinfo {author} {\bibfnamefont {J.}~\bibnamefont
			{Kono}}, \bibinfo {author} {\bibfnamefont {L.}~\bibnamefont {Wu}}, \bibinfo
		{author} {\bibfnamefont {K.~T.}\ \bibnamefont {Law}}, \bibinfo {author}
		{\bibfnamefont {E.}~\bibnamefont {Morosan}},\ and\ \bibinfo {author}
		{\bibfnamefont {M.}~\bibnamefont {Yi}},\ }\bibfield  {title} {\bibinfo
		{title} {Kramers nodal lines and weyl fermions in {SmAlSi}},\ }\bibfield
	{journal} {\bibinfo  {journal} {Communications Physics}\ }\textbf {\bibinfo
		{volume} {6}},\ \href {https://doi.org/10.1038/s42005-023-01257-2}
	{10.1038/s42005-023-01257-2} (\bibinfo {year} {2023})\BibitemShut {NoStop}%
	\bibitem [{\citenamefont {Ru}\ and\ \citenamefont {Fisher}(2006)}]{Ru2006}%
	\BibitemOpen
	\bibfield  {author} {\bibinfo {author} {\bibfnamefont {N.}~\bibnamefont
			{Ru}}\ and\ \bibinfo {author} {\bibfnamefont {I.~R.}\ \bibnamefont
			{Fisher}},\ }\bibfield  {title} {\bibinfo {title} {Thermodynamic and
			transport properties of {$\mathrm{Y}{\mathrm{Te}}_{3}$,
				$\mathrm{La}{\mathrm{Te}}_{3}$, and $\mathrm{Ce}{\mathrm{Te}}_{3}$}},\ }\href
	{https://doi.org/10.1103/PhysRevB.73.033101} {\bibfield  {journal} {\bibinfo
			{journal} {Phys. Rev. B}\ }\textbf {\bibinfo {volume} {73}},\ \bibinfo
		{pages} {033101} (\bibinfo {year} {2006})}\BibitemShut {NoStop}%
	\bibitem [{\citenamefont {Brouet}\ \emph {et~al.}(2008)\citenamefont {Brouet},
		\citenamefont {Yang}, \citenamefont {Zhou}, \citenamefont {Hussain},
		\citenamefont {Moore}, \citenamefont {He}, \citenamefont {Lu}, \citenamefont
		{Shen}, \citenamefont {Laverock}, \citenamefont {Dugdale}, \citenamefont
		{Ru},\ and\ \citenamefont {Fisher}}]{Brouet2008}%
	\BibitemOpen
	\bibfield  {author} {\bibinfo {author} {\bibfnamefont {V.}~\bibnamefont
			{Brouet}}, \bibinfo {author} {\bibfnamefont {W.~L.}\ \bibnamefont {Yang}},
		\bibinfo {author} {\bibfnamefont {X.~J.}\ \bibnamefont {Zhou}}, \bibinfo
		{author} {\bibfnamefont {Z.}~\bibnamefont {Hussain}}, \bibinfo {author}
		{\bibfnamefont {R.~G.}\ \bibnamefont {Moore}}, \bibinfo {author}
		{\bibfnamefont {R.}~\bibnamefont {He}}, \bibinfo {author} {\bibfnamefont
			{D.~H.}\ \bibnamefont {Lu}}, \bibinfo {author} {\bibfnamefont {Z.~X.}\
			\bibnamefont {Shen}}, \bibinfo {author} {\bibfnamefont {J.}~\bibnamefont
			{Laverock}}, \bibinfo {author} {\bibfnamefont {S.~B.}\ \bibnamefont
			{Dugdale}}, \bibinfo {author} {\bibfnamefont {N.}~\bibnamefont {Ru}},\ and\
		\bibinfo {author} {\bibfnamefont {I.~R.}\ \bibnamefont {Fisher}},\ }\bibfield
	{title} {\bibinfo {title} {Angle-resolved photoemission study of the
			evolution of band structure and charge density wave properties in {RTe}$_{3}$
			({R=Y}, {La, Ce, Sm, Gd, Tb, and Dy})},\ }\href
	{https://doi.org/10.1103/PhysRevB.77.235104} {\bibfield  {journal} {\bibinfo
			{journal} {Phys. Rev. B}\ }\textbf {\bibinfo {volume} {77}},\ \bibinfo
		{pages} {235104} (\bibinfo {year} {2008})}\BibitemShut {NoStop}%
	\bibitem [{\citenamefont {Yumigeta}\ \emph {et~al.}(2021)\citenamefont
		{Yumigeta}, \citenamefont {Qin}, \citenamefont {Li}, \citenamefont {Blei},
		\citenamefont {Attarde}, \citenamefont {Kopas},\ and\ \citenamefont
		{Tongay}}]{Yumigeta2021}%
	\BibitemOpen
	\bibfield  {author} {\bibinfo {author} {\bibfnamefont {K.}~\bibnamefont
			{Yumigeta}}, \bibinfo {author} {\bibfnamefont {Y.}~\bibnamefont {Qin}},
		\bibinfo {author} {\bibfnamefont {H.}~\bibnamefont {Li}}, \bibinfo {author}
		{\bibfnamefont {M.}~\bibnamefont {Blei}}, \bibinfo {author} {\bibfnamefont
			{Y.}~\bibnamefont {Attarde}}, \bibinfo {author} {\bibfnamefont
			{C.}~\bibnamefont {Kopas}},\ and\ \bibinfo {author} {\bibfnamefont
			{S.}~\bibnamefont {Tongay}},\ }\bibfield  {title} {\bibinfo {title} {Advances
			in rare-earth tritelluride quantum materials: Structure, properties, and
			synthesis},\ }\href {https://doi.org/10.1002/advs.202004762} {\bibfield
		{journal} {\bibinfo  {journal} {Adv. Sci.}\ }\textbf {\bibinfo {volume}
			{8}},\ \bibinfo {pages} {2004762} (\bibinfo {year} {2021})}\BibitemShut
	{NoStop}%
	\bibitem [{\citenamefont {Malliakas}\ and\ \citenamefont
		{Kanatzidis}(2006)}]{Malliakas2006}%
	\BibitemOpen
	\bibfield  {author} {\bibinfo {author} {\bibfnamefont {C.~D.}\ \bibnamefont
			{Malliakas}}\ and\ \bibinfo {author} {\bibfnamefont {M.~G.}\ \bibnamefont
			{Kanatzidis}},\ }\bibfield  {title} {\bibinfo {title} {Divergence in the
			behavior of the charge density wave in {RETe}$_3$ ({RE} = rare-earth element)
			with temperature and {RE} element},\ }\href
	{https://doi.org/10.1021/ja0641608} {\bibfield  {journal} {\bibinfo
			{journal} {J. Am. Chem. Soc.}\ }\textbf {\bibinfo {volume} {128}},\ \bibinfo
		{pages} {12612} (\bibinfo {year} {2006})}\BibitemShut {NoStop}%
	\bibitem [{\citenamefont {Ru}(2008)}]{RuThesis}%
	\BibitemOpen
	\bibfield  {author} {\bibinfo {author} {\bibfnamefont {N.}~\bibnamefont
			{Ru}},\ }\emph {\bibinfo {title} {Charge density wave formation in rare-earth
			tritellurides}},\ \href
	{https://doi.org/https://web.stanford.edu/group/fisher/people/Nancy_Ru_thesis.pdf}
	{Ph.D. thesis},\ \bibinfo  {school} {Stanford University}, \bibinfo {address}
	{California} (\bibinfo {year} {2008})\BibitemShut {NoStop}%
	\bibitem [{\citenamefont {He}\ \emph {et~al.}(2016)\citenamefont {He},
		\citenamefont {Wang}, \citenamefont {Yang}, \citenamefont {Long},
		\citenamefont {Zhao}, \citenamefont {Ma}, \citenamefont {Yang}, \citenamefont
		{Wang}, \citenamefont {Shangguan}, \citenamefont {Xue}, \citenamefont
		{Zhang}, \citenamefont {Ren}, \citenamefont {Li}, \citenamefont {Liu},\ and\
		\citenamefont {Chen}}]{He2016}%
	\BibitemOpen
	\bibfield  {author} {\bibinfo {author} {\bibfnamefont {J.~B.}\ \bibnamefont
			{He}}, \bibinfo {author} {\bibfnamefont {P.~P.}\ \bibnamefont {Wang}},
		\bibinfo {author} {\bibfnamefont {H.~X.}\ \bibnamefont {Yang}}, \bibinfo
		{author} {\bibfnamefont {Y.~J.}\ \bibnamefont {Long}}, \bibinfo {author}
		{\bibfnamefont {L.~X.}\ \bibnamefont {Zhao}}, \bibinfo {author}
		{\bibfnamefont {C.}~\bibnamefont {Ma}}, \bibinfo {author} {\bibfnamefont
			{M.}~\bibnamefont {Yang}}, \bibinfo {author} {\bibfnamefont {D.~M.}\
			\bibnamefont {Wang}}, \bibinfo {author} {\bibfnamefont {X.~C.}\ \bibnamefont
			{Shangguan}}, \bibinfo {author} {\bibfnamefont {M.~Q.}\ \bibnamefont {Xue}},
		\bibinfo {author} {\bibfnamefont {P.}~\bibnamefont {Zhang}}, \bibinfo
		{author} {\bibfnamefont {Z.~A.}\ \bibnamefont {Ren}}, \bibinfo {author}
		{\bibfnamefont {J.~Q.}\ \bibnamefont {Li}}, \bibinfo {author} {\bibfnamefont
			{W.~M.}\ \bibnamefont {Liu}},\ and\ \bibinfo {author} {\bibfnamefont {G.~F.}\
			\bibnamefont {Chen}},\ }\bibfield  {title} {\bibinfo {title}
		{Superconductivity in pd-intercalated charge-density-wave rare earth
			poly-tellurides rete$_n$},\ }\href
	{https://doi.org/10.1088/0953-2048/29/6/065018} {\bibfield  {journal}
		{\bibinfo  {journal} {Superconductor Science and Technology}\ }\textbf
		{\bibinfo {volume} {29}},\ \bibinfo {pages} {065018} (\bibinfo {year}
		{2016})}\BibitemShut {NoStop}%
	\bibitem [{\citenamefont {Pariari}\ \emph {et~al.}(2021)\citenamefont
		{Pariari}, \citenamefont {Koley}, \citenamefont {Roy}, \citenamefont
		{Singha}, \citenamefont {Laad}, \citenamefont {Taraphder},\ and\
		\citenamefont {Mandal}}]{Pariari2021}%
	\BibitemOpen
	\bibfield  {author} {\bibinfo {author} {\bibfnamefont {A.}~\bibnamefont
			{Pariari}}, \bibinfo {author} {\bibfnamefont {S.}~\bibnamefont {Koley}},
		\bibinfo {author} {\bibfnamefont {S.}~\bibnamefont {Roy}}, \bibinfo {author}
		{\bibfnamefont {R.}~\bibnamefont {Singha}}, \bibinfo {author} {\bibfnamefont
			{M.~S.}\ \bibnamefont {Laad}}, \bibinfo {author} {\bibfnamefont
			{A.}~\bibnamefont {Taraphder}},\ and\ \bibinfo {author} {\bibfnamefont
			{P.}~\bibnamefont {Mandal}},\ }\bibfield  {title} {\bibinfo {title}
		{Interplay between charge density wave order and magnetic field in the
			nonmagnetic rare-earth tritelluride {LaTe}$_{3}$},\ }\href
	{https://doi.org/10.1103/PhysRevB.104.155147} {\bibfield  {journal} {\bibinfo
			{journal} {Phys. Rev. B}\ }\textbf {\bibinfo {volume} {104}},\ \bibinfo
		{pages} {155147} (\bibinfo {year} {2021})}\BibitemShut {NoStop}%
	\bibitem [{\citenamefont {Petříček}\ \emph {et~al.}(2014)\citenamefont
		{Petříček}, \citenamefont {Dušek},\ and\ \citenamefont
		{Palatinus}}]{Petek2014}%
	\BibitemOpen
	\bibfield  {author} {\bibinfo {author} {\bibfnamefont {V.}~\bibnamefont
			{Petříček}}, \bibinfo {author} {\bibfnamefont {M.}~\bibnamefont
			{Dušek}},\ and\ \bibinfo {author} {\bibfnamefont {L.}~\bibnamefont
			{Palatinus}},\ }\bibfield  {title} {\bibinfo {title} {Crystallographic
			computing system jana2006: General features},\ }\href
	{https://doi.org/10.1515/zkri-2014-1737} {\bibfield  {journal} {\bibinfo
			{journal} {Zeitschrift f\"{u}r Kristallographie - Crystalline Materials}\
		}\textbf {\bibinfo {volume} {229}},\ \bibinfo {pages} {345–352} (\bibinfo
		{year} {2014})}\BibitemShut {NoStop}%
	\bibitem [{\citenamefont {Oszlányi}\ and\ \citenamefont
		{S\"{u}tő}(2004{\natexlab{a}})}]{Oszlnyi2004}%
	\BibitemOpen
	\bibfield  {author} {\bibinfo {author} {\bibfnamefont {G.}~\bibnamefont
			{Oszlányi}}\ and\ \bibinfo {author} {\bibfnamefont {A.}~\bibnamefont
			{S\"{u}tő}},\ }\bibfield  {title} {\bibinfo {title} {$ab~initio$~structure
			solution by charge flipping},\ }\href
	{https://doi.org/10.1107/s0108767303027569} {\bibfield  {journal} {\bibinfo
			{journal} {Acta Crystallographica Section A Foundations of Crystallography}\
		}\textbf {\bibinfo {volume} {60}},\ \bibinfo {pages} {134–141} (\bibinfo
		{year} {2004}{\natexlab{a}})}\BibitemShut {NoStop}%
	\bibitem [{\citenamefont {Oszlányi}\ and\ \citenamefont
		{S\"{u}tő}(2004{\natexlab{b}})}]{Oszlnyi2004_2}%
	\BibitemOpen
	\bibfield  {author} {\bibinfo {author} {\bibfnamefont {G.}~\bibnamefont
			{Oszlányi}}\ and\ \bibinfo {author} {\bibfnamefont {A.}~\bibnamefont
			{S\"{u}tő}},\ }\bibfield  {title} {\bibinfo {title} {Ab initio structure
			solution by charge flipping. ii. use of weak reflections},\ }\href
	{https://doi.org/10.1107/s0108767304027746} {\bibfield  {journal} {\bibinfo
			{journal} {Acta Crystallographica Section A Foundations of Crystallography}\
		}\textbf {\bibinfo {volume} {61}},\ \bibinfo {pages} {147–152} (\bibinfo
		{year} {2004}{\natexlab{b}})}\BibitemShut {NoStop}%
	\bibitem [{\citenamefont {Hoffmann}\ \emph {et~al.}(2004)\citenamefont
		{Hoffmann}, \citenamefont {S{\o}ndergaard}, \citenamefont {Schultz},
		\citenamefont {Li},\ and\ \citenamefont {Hofmann}}]{Hoffmann2004}%
	\BibitemOpen
	\bibfield  {author} {\bibinfo {author} {\bibfnamefont {S.}~\bibnamefont
			{Hoffmann}}, \bibinfo {author} {\bibfnamefont {C.}~\bibnamefont
			{S{\o}ndergaard}}, \bibinfo {author} {\bibfnamefont {C.}~\bibnamefont
			{Schultz}}, \bibinfo {author} {\bibfnamefont {Z.}~\bibnamefont {Li}},\ and\
		\bibinfo {author} {\bibfnamefont {P.}~\bibnamefont {Hofmann}},\ }\bibfield
	{title} {\bibinfo {title} {An undulator-based spherical grating monochromator
			beamline for angle-resolved photoemission spectroscopy},\ }\href
	{https://doi.org/10.1016/j.nima.2004.01.039} {\bibfield  {journal} {\bibinfo
			{journal} {Nucl. Instrum. Methods Phys. Res. A}\ }\textbf {\bibinfo {volume}
			{523}},\ \bibinfo {pages} {441} (\bibinfo {year} {2004})}\BibitemShut
	{NoStop}%
	\bibitem [{Igo(2021)}]{IgorProManualV9}%
	\BibitemOpen
	\href@noop {} {\emph {\bibinfo {title} {Igor Pro Manual}}},\ \bibinfo
	{edition} {version 9}\ ed.\ (\bibinfo {year} {2021})\BibitemShut {NoStop}%
	\bibitem [{\citenamefont {Kresse}\ and\ \citenamefont
		{Furthm\"uller}(1996)}]{Kresse_1996}%
	\BibitemOpen
	\bibfield  {author} {\bibinfo {author} {\bibfnamefont {G.}~\bibnamefont
			{Kresse}}\ and\ \bibinfo {author} {\bibfnamefont {J.}~\bibnamefont
			{Furthm\"uller}},\ }\bibfield  {title} {\bibinfo {title} {Efficient iterative
			schemes for $ab-initio$ total-energy calculations using a plane-wave basis
			set},\ }\href {https://doi.org/10.1103/PhysRevB.54.11169} {\bibfield
		{journal} {\bibinfo  {journal} {Phys. Rev. B}\ }\textbf {\bibinfo {volume}
			{54}},\ \bibinfo {pages} {11169} (\bibinfo {year} {1996})}\BibitemShut
	{NoStop}%
	\bibitem [{\citenamefont {Kresse}\ and\ \citenamefont
		{Joubert}(1999)}]{Kresse_1999}%
	\BibitemOpen
	\bibfield  {author} {\bibinfo {author} {\bibfnamefont {G.}~\bibnamefont
			{Kresse}}\ and\ \bibinfo {author} {\bibfnamefont {D.}~\bibnamefont
			{Joubert}},\ }\bibfield  {title} {\bibinfo {title} {From ultrasoft
			pseudopotentials to the projector augmented-wave method},\ }\href
	{https://doi.org/10.1103/PhysRevB.59.1758} {\bibfield  {journal} {\bibinfo
			{journal} {Phys. Rev. B}\ }\textbf {\bibinfo {volume} {59}},\ \bibinfo
		{pages} {1758} (\bibinfo {year} {1999})}\BibitemShut {NoStop}%
	\bibitem [{\citenamefont {Perdew}\ \emph {et~al.}(1996)\citenamefont {Perdew},
		\citenamefont {Burke},\ and\ \citenamefont {Ernzerhof}}]{Perdew}%
	\BibitemOpen
	\bibfield  {author} {\bibinfo {author} {\bibfnamefont {J.~P.}\ \bibnamefont
			{Perdew}}, \bibinfo {author} {\bibfnamefont {K.}~\bibnamefont {Burke}},\ and\
		\bibinfo {author} {\bibfnamefont {M.}~\bibnamefont {Ernzerhof}},\ }\bibfield
	{title} {\bibinfo {title} {Generalized gradient approximation made simple},\
	}\href {https://doi.org/10.1103/PhysRevLett.77.3865} {\bibfield  {journal}
		{\bibinfo  {journal} {Phys. Rev. Lett.}\ }\textbf {\bibinfo {volume} {77}},\
		\bibinfo {pages} {3865} (\bibinfo {year} {1996})}\BibitemShut {NoStop}%
	\bibitem [{\citenamefont {Capillas}\ \emph {et~al.}(2011)\citenamefont
		{Capillas}, \citenamefont {Tasci}, \citenamefont {de~la Flor}, \citenamefont
		{Orobengoa}, \citenamefont {Perez-Mato},\ and\ \citenamefont
		{Aroyo}}]{Capillas2011}%
	\BibitemOpen
	\bibfield  {author} {\bibinfo {author} {\bibfnamefont {C.}~\bibnamefont
			{Capillas}}, \bibinfo {author} {\bibfnamefont {E.~S.}\ \bibnamefont {Tasci}},
		\bibinfo {author} {\bibfnamefont {G.}~\bibnamefont {de~la Flor}}, \bibinfo
		{author} {\bibfnamefont {D.}~\bibnamefont {Orobengoa}}, \bibinfo {author}
		{\bibfnamefont {J.~M.}\ \bibnamefont {Perez-Mato}},\ and\ \bibinfo {author}
		{\bibfnamefont {M.~I.}\ \bibnamefont {Aroyo}},\ }\bibfield  {title} {\bibinfo
		{title} {A new computer tool at the bilbao crystallographic server to detect
			and characterize pseudosymmetry},\ }\href
	{https://doi.org/10.1524/zkri.2011.1321} {\bibfield  {journal} {\bibinfo
			{journal} {Z. Kristallogr.}\ }\textbf {\bibinfo {volume} {226}},\ \bibinfo
		{pages} {186} (\bibinfo {year} {2011})}\BibitemShut {NoStop}%
	\bibitem [{\citenamefont {Herath}\ \emph {et~al.}(2020)\citenamefont {Herath},
		\citenamefont {Tavadze}, \citenamefont {He}, \citenamefont {Bousquet},
		\citenamefont {Singh}, \citenamefont {Mu{\~{n}}oz},\ and\ \citenamefont
		{Romero}}]{Herath2020}%
	\BibitemOpen
	\bibfield  {author} {\bibinfo {author} {\bibfnamefont {U.}~\bibnamefont
			{Herath}}, \bibinfo {author} {\bibfnamefont {P.}~\bibnamefont {Tavadze}},
		\bibinfo {author} {\bibfnamefont {X.}~\bibnamefont {He}}, \bibinfo {author}
		{\bibfnamefont {E.}~\bibnamefont {Bousquet}}, \bibinfo {author}
		{\bibfnamefont {S.}~\bibnamefont {Singh}}, \bibinfo {author} {\bibfnamefont
			{F.}~\bibnamefont {Mu{\~{n}}oz}},\ and\ \bibinfo {author} {\bibfnamefont
			{A.~H.}\ \bibnamefont {Romero}},\ }\bibfield  {title} {\bibinfo {title}
		{{PyProcar}: A python library for electronic structure pre/post-processing},\
	}\href {https://doi.org/10.1016/j.cpc.2019.107080} {\bibfield  {journal}
		{\bibinfo  {journal} {Comput. Phys. Commun.}\ }\textbf {\bibinfo {volume}
			{251}},\ \bibinfo {pages} {107080} (\bibinfo {year} {2020})}\BibitemShut
	{NoStop}%
	\bibitem [{\citenamefont {Momma}\ and\ \citenamefont
		{Izumi}(2011)}]{Momma2011}%
	\BibitemOpen
	\bibfield  {author} {\bibinfo {author} {\bibfnamefont {K.}~\bibnamefont
			{Momma}}\ and\ \bibinfo {author} {\bibfnamefont {F.}~\bibnamefont {Izumi}},\
	}\bibfield  {title} {\bibinfo {title} {Vesta $3$ for three-dimensional
			visualization of crystal, volumetric and morphology data},\ }\href
	{https://doi.org/10.1107/s0021889811038970} {\bibfield  {journal} {\bibinfo
			{journal} {J. Appl. Crystallogr.}\ }\textbf {\bibinfo {volume} {44}},\
		\bibinfo {pages} {1272} (\bibinfo {year} {2011})}\BibitemShut {NoStop}%
	\bibitem [{sup()}]{supp}%
	\BibitemOpen
	\bibinfo {note} {See Supplemental Material at xx.xx for Figs. S1-S5, Table S1
		{and} Discussion A}\BibitemShut {NoStop}%
	\bibitem [{\citenamefont {Janssen}\ \emph {et~al.}(2006)\citenamefont
		{Janssen}, \citenamefont {Janner}, \citenamefont {Looijenga-Vos},\ and\
		\citenamefont {de~Wolff}}]{Janssen2006}%
	\BibitemOpen
	\bibfield  {author} {\bibinfo {author} {\bibfnamefont {T.}~\bibnamefont
			{Janssen}}, \bibinfo {author} {\bibfnamefont {A.}~\bibnamefont {Janner}},
		\bibinfo {author} {\bibfnamefont {A.}~\bibnamefont {Looijenga-Vos}},\ and\
		\bibinfo {author} {\bibfnamefont {P.~M.}\ \bibnamefont {de~Wolff}},\
	}\bibfield  {title} {\bibinfo {title} {Incommensurate and commensurate
			modulated structures},\ }in\ \href
	{https://doi.org/10.1107/97809553602060000624} {\emph {\bibinfo {booktitle}
			{International Tables for Crystallography}}}\ (\bibinfo  {publisher}
	{International Union of Crystallography},\ \bibinfo {year} {2006})\ pp.\
	\bibinfo {pages} {907--955}\BibitemShut {NoStop}%
	\bibitem [{\citenamefont {Van~Smaalen}(2007)}]{van2007incommensurate}%
	\BibitemOpen
	\bibfield  {author} {\bibinfo {author} {\bibfnamefont {S.}~\bibnamefont
			{Van~Smaalen}},\ }\href@noop {} {\emph {\bibinfo {title} {Incommensurate
				crystallography}}},\ Vol.~\bibinfo {volume} {21}\ (\bibinfo  {publisher}
	{Oxford University Press},\ \bibinfo {year} {2007})\BibitemShut {NoStop}%
	\bibitem [{\citenamefont {Wyman}\ and\ \citenamefont
		{Wyman}(1985)}]{Wyman1985}%
	\BibitemOpen
	\bibfield  {author} {\bibinfo {author} {\bibfnamefont {M.~F.}\ \bibnamefont
			{Wyman}}\ and\ \bibinfo {author} {\bibfnamefont {B.~F.}\ \bibnamefont
			{Wyman}},\ }\bibfield  {title} {\bibinfo {title} {An essay on continued
			fractions},\ }\href {https://doi.org/10.1007/bf01699475} {\bibfield
		{journal} {\bibinfo  {journal} {Math. Syst. Theory}\ }\textbf {\bibinfo
			{volume} {18}},\ \bibinfo {pages} {295} (\bibinfo {year} {1985})}\BibitemShut
	{NoStop}%
	\bibitem [{\citenamefont {Sarkar}\ \emph {et~al.}(2020)\citenamefont {Sarkar},
		\citenamefont {Singh}, \citenamefont {Sadhukhan}, \citenamefont {Pariari},
		\citenamefont {Roy}, \citenamefont {Mandal},\ and\ \citenamefont
		{Barman}}]{SarkarAIP2020}%
	\BibitemOpen
	\bibfield  {author} {\bibinfo {author} {\bibfnamefont {S.}~\bibnamefont
			{Sarkar}}, \bibinfo {author} {\bibfnamefont {V.~K.}\ \bibnamefont {Singh}},
		\bibinfo {author} {\bibfnamefont {P.}~\bibnamefont {Sadhukhan}}, \bibinfo
		{author} {\bibfnamefont {A.}~\bibnamefont {Pariari}}, \bibinfo {author}
		{\bibfnamefont {S.}~\bibnamefont {Roy}}, \bibinfo {author} {\bibfnamefont
			{P.}~\bibnamefont {Mandal}},\ and\ \bibinfo {author} {\bibfnamefont {S.~R.}\
			\bibnamefont {Barman}},\ }\bibfield  {title} {\bibinfo {title} {X-ray
			photoelectron spectroscopy study of a layered tri-chalcogenide system
			{LaTe}$_3$},\ }\href {https://doi.org/10.1063/5.0001764} {\bibfield
		{journal} {\bibinfo  {journal} {{AIP} Conf. Proc.}\ }\textbf {\bibinfo
			{volume} {2220}},\ \bibinfo {pages} {100005} (\bibinfo {year}
		{2020})}\BibitemShut {NoStop}%
	\bibitem [{\citenamefont {Ru}\ \emph {et~al.}(2008)\citenamefont {Ru},
		\citenamefont {Condron}, \citenamefont {Margulis}, \citenamefont {Shin},
		\citenamefont {Laverock}, \citenamefont {Dugdale}, \citenamefont {Toney},\
		and\ \citenamefont {Fisher}}]{RuPRB2008}%
	\BibitemOpen
	\bibfield  {author} {\bibinfo {author} {\bibfnamefont {N.}~\bibnamefont
			{Ru}}, \bibinfo {author} {\bibfnamefont {C.~L.}\ \bibnamefont {Condron}},
		\bibinfo {author} {\bibfnamefont {G.~Y.}\ \bibnamefont {Margulis}}, \bibinfo
		{author} {\bibfnamefont {K.~Y.}\ \bibnamefont {Shin}}, \bibinfo {author}
		{\bibfnamefont {J.}~\bibnamefont {Laverock}}, \bibinfo {author}
		{\bibfnamefont {S.~B.}\ \bibnamefont {Dugdale}}, \bibinfo {author}
		{\bibfnamefont {M.~F.}\ \bibnamefont {Toney}},\ and\ \bibinfo {author}
		{\bibfnamefont {I.~R.}\ \bibnamefont {Fisher}},\ }\bibfield  {title}
	{\bibinfo {title} {Effect of chemical pressure on the charge density wave
			transition in rare-earth tritellurides {RTe}$_3$},\ }\href
	{https://doi.org/10.1103/physrevb.77.035114} {\bibfield  {journal} {\bibinfo
			{journal} {Phys. Rev. B}\ }\textbf {\bibinfo {volume} {77}},\ \bibinfo
		{pages} {035114} (\bibinfo {year} {2008})}\BibitemShut {NoStop}%
	\bibitem [{\citenamefont {Fang}\ \emph {et~al.}(2007)\citenamefont {Fang},
		\citenamefont {Ru}, \citenamefont {Fisher},\ and\ \citenamefont
		{Kapitulnik}}]{Fang2007}%
	\BibitemOpen
	\bibfield  {author} {\bibinfo {author} {\bibfnamefont {A.}~\bibnamefont
			{Fang}}, \bibinfo {author} {\bibfnamefont {N.}~\bibnamefont {Ru}}, \bibinfo
		{author} {\bibfnamefont {I.~R.}\ \bibnamefont {Fisher}},\ and\ \bibinfo
		{author} {\bibfnamefont {A.}~\bibnamefont {Kapitulnik}},\ }\bibfield  {title}
	{\bibinfo {title} {Stm studies of ${\mathrm{tbte}}_{3}$: Evidence for a fully
			incommensurate charge density wave},\ }\href
	{https://doi.org/10.1103/PhysRevLett.99.046401} {\bibfield  {journal}
		{\bibinfo  {journal} {Phys. Rev. Lett.}\ }\textbf {\bibinfo {volume} {99}},\
		\bibinfo {pages} {046401} (\bibinfo {year} {2007})}\BibitemShut {NoStop}%
	\bibitem [{\citenamefont {Moore}\ \emph {et~al.}(2010)\citenamefont {Moore},
		\citenamefont {Brouet}, \citenamefont {He}, \citenamefont {Lu}, \citenamefont
		{Ru}, \citenamefont {Chu}, \citenamefont {Fisher},\ and\ \citenamefont
		{Shen}}]{Moore2010}%
	\BibitemOpen
	\bibfield  {author} {\bibinfo {author} {\bibfnamefont {R.~G.}\ \bibnamefont
			{Moore}}, \bibinfo {author} {\bibfnamefont {V.}~\bibnamefont {Brouet}},
		\bibinfo {author} {\bibfnamefont {R.}~\bibnamefont {He}}, \bibinfo {author}
		{\bibfnamefont {D.~H.}\ \bibnamefont {Lu}}, \bibinfo {author} {\bibfnamefont
			{N.}~\bibnamefont {Ru}}, \bibinfo {author} {\bibfnamefont {J.-H.}\
			\bibnamefont {Chu}}, \bibinfo {author} {\bibfnamefont {I.~R.}\ \bibnamefont
			{Fisher}},\ and\ \bibinfo {author} {\bibfnamefont {Z.-X.}\ \bibnamefont
			{Shen}},\ }\bibfield  {title} {\bibinfo {title} {Fermi surface evolution
			across multiple charge density wave transitions in ${\text{erte}}_{3}$},\
	}\href {https://doi.org/10.1103/PhysRevB.81.073102} {\bibfield  {journal}
		{\bibinfo  {journal} {Phys. Rev. B}\ }\textbf {\bibinfo {volume} {81}},\
		\bibinfo {pages} {073102} (\bibinfo {year} {2010})}\BibitemShut {NoStop}%
	\bibitem [{\citenamefont {Lavagnini}\ \emph {et~al.}(2008)\citenamefont
		{Lavagnini}, \citenamefont {Baldini}, \citenamefont {Sacchetti},
		\citenamefont {Di~Castro}, \citenamefont {Delley}, \citenamefont {Monnier},
		\citenamefont {Chu}, \citenamefont {Ru}, \citenamefont {Fisher},
		\citenamefont {Postorino},\ and\ \citenamefont {Degiorgi}}]{Lavagnini2008}%
	\BibitemOpen
	\bibfield  {author} {\bibinfo {author} {\bibfnamefont {M.}~\bibnamefont
			{Lavagnini}}, \bibinfo {author} {\bibfnamefont {M.}~\bibnamefont {Baldini}},
		\bibinfo {author} {\bibfnamefont {A.}~\bibnamefont {Sacchetti}}, \bibinfo
		{author} {\bibfnamefont {D.}~\bibnamefont {Di~Castro}}, \bibinfo {author}
		{\bibfnamefont {B.}~\bibnamefont {Delley}}, \bibinfo {author} {\bibfnamefont
			{R.}~\bibnamefont {Monnier}}, \bibinfo {author} {\bibfnamefont {J.-H.}\
			\bibnamefont {Chu}}, \bibinfo {author} {\bibfnamefont {N.}~\bibnamefont
			{Ru}}, \bibinfo {author} {\bibfnamefont {I.~R.}\ \bibnamefont {Fisher}},
		\bibinfo {author} {\bibfnamefont {P.}~\bibnamefont {Postorino}},\ and\
		\bibinfo {author} {\bibfnamefont {L.}~\bibnamefont {Degiorgi}},\ }\bibfield
	{title} {\bibinfo {title} {Evidence for coupling between charge density waves
			and phonons in two-dimensional rare-earth tritellurides},\ }\href
	{https://doi.org/10.1103/PhysRevB.78.201101} {\bibfield  {journal} {\bibinfo
			{journal} {Phys. Rev. B}\ }\textbf {\bibinfo {volume} {78}},\ \bibinfo
		{pages} {201101} (\bibinfo {year} {2008})}\BibitemShut {NoStop}%
	\bibitem [{\citenamefont {Setyawan}\ and\ \citenamefont
		{Curtarolo}(2010)}]{Setyawan2010}%
	\BibitemOpen
	\bibfield  {author} {\bibinfo {author} {\bibfnamefont {W.}~\bibnamefont
			{Setyawan}}\ and\ \bibinfo {author} {\bibfnamefont {S.}~\bibnamefont
			{Curtarolo}},\ }\bibfield  {title} {\bibinfo {title} {High-throughput
			electronic band structure calculations: Challenges and tools},\ }\href
	{https://doi.org/10.1016/j.commatsci.2010.05.010} {\bibfield  {journal}
		{\bibinfo  {journal} {Comput. Mater. Sci.}\ }\textbf {\bibinfo {volume}
			{49}},\ \bibinfo {pages} {299} (\bibinfo {year} {2010})}\BibitemShut
	{NoStop}%
	\bibitem [{\citenamefont {Ku}\ \emph {et~al.}(2010)\citenamefont {Ku},
		\citenamefont {Berlijn},\ and\ \citenamefont {Lee}}]{Ku2010}%
	\BibitemOpen
	\bibfield  {author} {\bibinfo {author} {\bibfnamefont {W.}~\bibnamefont
			{Ku}}, \bibinfo {author} {\bibfnamefont {T.}~\bibnamefont {Berlijn}},\ and\
		\bibinfo {author} {\bibfnamefont {C.-C.}\ \bibnamefont {Lee}},\ }\bibfield
	{title} {\bibinfo {title} {Unfolding first-principles band structures},\
	}\href {https://doi.org/10.1103/PhysRevLett.104.216401} {\bibfield  {journal}
		{\bibinfo  {journal} {Phys. Rev. Lett.}\ }\textbf {\bibinfo {volume} {104}},\
		\bibinfo {pages} {216401} (\bibinfo {year} {2010})}\BibitemShut {NoStop}%
	\bibitem [{\citenamefont {Allen}\ \emph {et~al.}(2013)\citenamefont {Allen},
		\citenamefont {Berlijn}, \citenamefont {Casavant},\ and\ \citenamefont
		{Soler}}]{Allen2013}%
	\BibitemOpen
	\bibfield  {author} {\bibinfo {author} {\bibfnamefont {P.~B.}\ \bibnamefont
			{Allen}}, \bibinfo {author} {\bibfnamefont {T.}~\bibnamefont {Berlijn}},
		\bibinfo {author} {\bibfnamefont {D.~A.}\ \bibnamefont {Casavant}},\ and\
		\bibinfo {author} {\bibfnamefont {J.~M.}\ \bibnamefont {Soler}},\ }\bibfield
	{title} {\bibinfo {title} {Recovering hidden bloch character: Unfolding
			electrons, phonons, and slabs},\ }\href
	{https://doi.org/10.1103/PhysRevB.87.085322} {\bibfield  {journal} {\bibinfo
			{journal} {Phys. Rev. B}\ }\textbf {\bibinfo {volume} {87}},\ \bibinfo
		{pages} {085322} (\bibinfo {year} {2013})}\BibitemShut {NoStop}%
	\bibitem [{\citenamefont {Gweon}\ \emph {et~al.}(1998)\citenamefont {Gweon},
		\citenamefont {Denlinger}, \citenamefont {Clack}, \citenamefont {Allen},
		\citenamefont {Olson}, \citenamefont {DiMasi}, \citenamefont {Aronson},
		\citenamefont {Foran},\ and\ \citenamefont {Lee}}]{Gweon1998}%
	\BibitemOpen
	\bibfield  {author} {\bibinfo {author} {\bibfnamefont {G.-H.}\ \bibnamefont
			{Gweon}}, \bibinfo {author} {\bibfnamefont {J.~D.}\ \bibnamefont
			{Denlinger}}, \bibinfo {author} {\bibfnamefont {J.~A.}\ \bibnamefont
			{Clack}}, \bibinfo {author} {\bibfnamefont {J.~W.}\ \bibnamefont {Allen}},
		\bibinfo {author} {\bibfnamefont {C.~G.}\ \bibnamefont {Olson}}, \bibinfo
		{author} {\bibfnamefont {E.}~\bibnamefont {DiMasi}}, \bibinfo {author}
		{\bibfnamefont {M.~C.}\ \bibnamefont {Aronson}}, \bibinfo {author}
		{\bibfnamefont {B.}~\bibnamefont {Foran}},\ and\ \bibinfo {author}
		{\bibfnamefont {S.}~\bibnamefont {Lee}},\ }\bibfield  {title} {\bibinfo
		{title} {Direct observation of complete fermi surface, imperfect nesting, and
			gap anisotropy in the high-temperature incommensurate charge-density-wave
			compound {SmTe}$_{3}$},\ }\href {https://doi.org/10.1103/PhysRevLett.81.886}
	{\bibfield  {journal} {\bibinfo  {journal} {Phys. Rev. Lett.}\ }\textbf
		{\bibinfo {volume} {81}},\ \bibinfo {pages} {886} (\bibinfo {year}
		{1998})}\BibitemShut {NoStop}%
	\bibitem [{\citenamefont {Chikina}\ \emph {et~al.}(2023)\citenamefont
		{Chikina}, \citenamefont {Lund}, \citenamefont {Bianchi}, \citenamefont
		{Curcio}, \citenamefont {Dalgaard}, \citenamefont {Bremholm}, \citenamefont
		{Lei}, \citenamefont {Singha}, \citenamefont {Schoop},\ and\ \citenamefont
		{Hofmann}}]{Chikina2023}%
	\BibitemOpen
	\bibfield  {author} {\bibinfo {author} {\bibfnamefont {A.}~\bibnamefont
			{Chikina}}, \bibinfo {author} {\bibfnamefont {H.}~\bibnamefont {Lund}},
		\bibinfo {author} {\bibfnamefont {M.}~\bibnamefont {Bianchi}}, \bibinfo
		{author} {\bibfnamefont {D.}~\bibnamefont {Curcio}}, \bibinfo {author}
		{\bibfnamefont {K.~J.}\ \bibnamefont {Dalgaard}}, \bibinfo {author}
		{\bibfnamefont {M.}~\bibnamefont {Bremholm}}, \bibinfo {author}
		{\bibfnamefont {S.}~\bibnamefont {Lei}}, \bibinfo {author} {\bibfnamefont
			{R.}~\bibnamefont {Singha}}, \bibinfo {author} {\bibfnamefont {L.~M.}\
			\bibnamefont {Schoop}},\ and\ \bibinfo {author} {\bibfnamefont
			{P.}~\bibnamefont {Hofmann}},\ }\bibfield  {title} {\bibinfo {title} {Charge
			density wave generated fermi surfaces in ${\mathrm{ndte}}_{3}$},\ }\href
	{https://doi.org/10.1103/PhysRevB.107.L161103} {\bibfield  {journal}
		{\bibinfo  {journal} {Phys. Rev. B}\ }\textbf {\bibinfo {volume} {107}},\
		\bibinfo {pages} {L161103} (\bibinfo {year} {2023})}\BibitemShut {NoStop}%
\end{thebibliography}
%
\end{document}